\documentclass[12pt]{article}
\usepackage{graphicx}
\usepackage{amsmath}
\newcommand{\sm}{\hbox{$\bigcirc$\kern-0.72em\hbox{\bf s} }}
\newcommand{\Id}{\hbox{\sl 1\kern-0.25em\hbox{I}}}

\newcommand{\rcorr}{\hbox{\kern-1.2em$\longrightarrow$}}
\newcommand{\lrcorr}{\hbox{\kern-1.2em$\longleftrightarrow$}}

\newcommand{\nRightarrow}{\Rightarrow\kern-1.2em\hbox{/}\kern.8em} %
\newcommand{\BB}{\hbox{I\kern-.2em\hbox{B}}} 
\newcommand{\DD}{\hbox{I\kern-.2em\hbox{D}}} 
\newcommand{\FF}{\hbox{I\kern-.2em\hbox{F}}} 
\newcommand{\NN}{\hbox{I\kern-.2em\hbox{N}}}  
\newcommand{\ZZ}{{{\rm Z}\kern-.28em{\rm Z}}} 
\newcommand{\RR}{\mathop{{\rm I}\kern-.2em{\rm R}}\nolimits} 
\newcommand{\RRe}{\mathop{{\rm I}\kern-.2em{\rm Re}}\nolimits} 
\newcommand{\QQ}{\hbox{l\kern-.36em\hbox{Q}}}  
\newcommand{\CC}{\hbox{{\textsf I}\kern-.47em\hbox{C}}}
\newcommand{\nop}{\hbox{{\textsf I}\kern-.47em\hbox{O}}}

\begin{document}
\title{Group theoretical foundations of the\\ Quantum Theory of an interacting particle}
\author{Giuseppe Nistic\`o
\\
{\small Dipartimento
di Matematica e Informatica, Universit\`a della Calabria, Italy}\\
{\small and}\\
{\small
INFN -- gruppo collegato di Cosenza, Italy}\\
{\small email: gnistico@unical.it} } \maketitle
\abstract{We point out some obstacles raised by the lost of symmetry against the
extension to the case of an interacting particle of the approach
that {\sl deductively} establishes the Quantum Theory of a free particle
according to the group theoretical methods worked out by Bargmann, Mackey and Wigner.
Then we develop an approach
which overcomes these difficulties in the non relativistic case.
According to our approach the different specific forms of the
wave equation of an interacting particle are implied by
particular first order invariance properties that characterize the interaction
with respect to specific sub-groups
of galileian transformations.
Moreover, the possibility of yet unknown forms of the wave equation is left open.}
\section{Introduction}
It is well known that
group theoretical methods, due in particular to E. Wigner and G. Mackey,
allow to attain a formulation of the Quantum Theory of a free particle through
a purely deductive development based on symmetry principles.
These approaches enforce the circumstance that Galilei's group $\mathcal G$
(or Poincar\'e's group $\mathcal P$, for a relativistic theory) is a group
of symmetry transformations for an isolated particle, so that Wigner's
theorem \cite{b1},\cite{b2} on the representation of symmetries and
Mackey's imprimitivity theorem \cite{b3},\cite{b4} can
be applied to deduce the explicit Quantum Theory of a free particle \cite{b5}.
In so doing, it is avoided  canonical quantization, which
invokes a pre-existing {\sl classical theory} of the physical system under investigation and formulates
its Quantum Theory  by  replacing classical magnitudes with operators.
\par
The extension of these group theoretical methods,
so satisfactory for a free particle, to an interacting particle encounters serious
problems; the  main obstacle is the fact that for a non-isolated
system the galileian transformations, or the transformations of Poincar\'e in the relativistic case,
do not form a group of symmetry transformations \cite{c6},
so that neither Wigner's theorem nor Mackey's
imprimitivity theorem  can directly apply.
\par
In fact, in the literature several approaches can be found that extend the group theoretical methods to
the interacting particle case. However, all these approaches \cite{b5},\cite{c7},\cite{c8},\cite{c6}, in order to
overcome the difficulty raised by the lost of symmetry, have to introduce some assumptions; but, as
we argue in section 2.4, these assumptions lead to an empirically non adequate theory,
unable, in particular, to describe particles interacting with electromagnetic fields.
\par
In this work we show how a group theoretical approach to the Quantum Theory of an interacting particle
can be successfully pursued without introducing assumptions as those required in
\cite{b5},\cite{c7},\cite{c8},\cite{c6}, that too severely restrict the empirical domain of the theory.
The present article addresses this task in the non-relativistic case;
the development of the approach for a relativistic theory is in progress.
\vskip.5pc
First we find preliminary  results which hold both for a non-relativistic and for a relativistic theory, i.e.
independently of which group $\Upsilon$ of space-time transformations, $\Upsilon=\mathcal G$ or $\Upsilon=\mathcal P$,
is taken into account.
The basic concept (QT) of {\sl quantum transformation} corresponding to a space-time transformation $g\in\Upsilon$, viable
also in absence of the condition of symmetry, is introduced in section 2.2 as a transformation $S_g^\Sigma$ defined on the whole set
of quantum observables,
which  in general depends also on the reference frame $\Sigma$.
The  conditions to be required to this
more general notion of quantum transformation are identified in sections 2.2 and 3.1 as three constraints (S.1), (S.2), (S.3).
We show in sections 3.1, 3.2 that these conditions, together with a continuity condition for $g\to S_g^\Sigma$,
imply that every transformation $g\in\Upsilon$ can be assigned a {\sl unitary} operator $U_g$,
also if the $g$'s are not symmetries, that realizes the quantum transformation
of a quantum observable $A$ as $S_g^\Sigma[A]=U_gAU_g^{-1}$, in such a way that
the correspondence $g\to U_g$ is a continuous mapping.
\par
But at this point an obstacle stops our development, because the properties (S.1), (S.2), (S.3) are not sufficient to imply that
$g\to U_g$ is a {\sl projective representation}, that is one of the conditions required in order that
the imprimitivity theorem applies to proceed in the approach.
\par
To address this problem, in section 3.3
we introduce the notion of $\sigma$-conversion, which is a straight mathematical procedure
that converts each $U_g$ into another unitary operator $\hat U_g$ in such a way that
$g\to\hat U_g$ is a {\sl projective representation}.
In the non relativistic case, we prove that the imprimitivity theorem for the {\sl euclidean}
group $\mathcal E$
-- not for the whole galileian group $\mathcal G$ --
can be applied to explicitly identify a mathematical formalism of the theory;
but in general the position operators are not explicitly identified, so that
the identified formalism turns out to be devoid of physical significance.
\par
In order to attain an effective theory it is necessary to determine
which operators represent position and to determine the dynamical law.
In section 3.4 we show how the operators that {\sl physically} represent the position of the particle
are explicitly represented for a particular class of interactions, completely characterized by admitting
``Q-covariant'' $\sigma$-conversions, i.e. $\sigma$-conversions
that leaves unaltered the covariance properties of the position with respect to $\mathcal G$.
For this class of interactions the general dynamical law is determined in section 3.5.
\par
This general dynamical law does not specify the explicit form of the hamiltonian operator $H$;
in fact, different specific forms of the wave equation are compatible with this general law. Then we face the
problem of singling out conditions related to the interaction, which determine the different specific
wave equations.
\par
In section 4 we identify these conditions as {\sl invariance properties related to the interaction}.
More precisely, we single out which {\sl specific forms the wave equation takes} if the $\sigma$-conversion
admitted by the interaction leaves unaltered, at the first order, the covariance properties of ${\bf Q}^{(t)}$,
namely of position at time $t$, {\sl with respect to specific sub-groups of $\mathcal G$}.
In so doing, the known wave equations are recovered, but also yet unknown ones could be singled out.
In the conclusive section 4.4 the relation of the present approach with other
methods for quantizing the interaction are briefly discussed.
\section{Space-time and quantum transformations}
In this section, once introduced some necessary mathematics, we establish the concepts
that enforce the work and formulate them in the quantum formalism.
In section 2.2 we introduce a concept of {\sl quantum transformation},
corresponding to  Galilei's or Poincar\'e's transformations,
that is viable also in the case that the system is an {\sl interacting}
system, i.e., when the transformations are not  symmetries.
A general property (S.1) of these quantum transformations, conceptually
entailed by their very meaning, is identified.
The presence of the {\sl symmetry} character of the transformations implies stronger properties; they are established in
section 2.3, where we outline how the further properties can be used to obtain the explicit Quantum Theory  of a free particle
by mathematically deducing it form the symmetry principles.
This outline allow us to identify, in sections 2.4, the obstacles raised by the lost of the symmetry condition
 against a similar deduction in the case of an interacting particle.
\subsection{Mathematical tools}
Let us introduce, to begin, the notation for the mathematical structures involved in the work.
The Quantum Theory of a
physical system, formulated in a complex and separable Hilbert space $\mathcal H$, needs the following mathematical structures.
\begin{description}
\item[\quad-]
The set $\Omega({\mathcal H})$ of all self-adjoint operators of $\mathcal H$, which represent
{\sl quantum observables}.
\item[\quad-]
The complete, ortho-complemented lattice $\Pi({\mathcal H})$ of all projections operators of $\mathcal H$, i.e.
quantum observables with possible outcomes
in $\{0,1\}$.
\item[\quad-]
The set $\Pi_1({\mathcal H})$ of all rank one orthogonal projections of $\mathcal H$.
\item[\quad-]
The set ${\mathcal S}({\mathcal H})$ of all density operators of $\mathcal H$,
which represent {\sl quantum states}.
\item[\quad-]
The set ${\mathcal U}({\mathcal H})$ of all unitary operators of the Hilbert space $\mathcal H$.
\end{description}
In the group theoretical approach a key role is played by the  {\sl imprimitivity theorem}
of Mackey, which is a representation theorem for {\sl imprimitivity} systems
relative to projective representations \cite{b4}.
The following definition recalls the notion of projective representation.
\vskip.5pc\noindent
{\bf Definition 2.1.} {\sl Let $G$ be a separable, locally compact group with identity element $e$. A correspondence
$U:G\to{\mathcal U}({\mathcal H})$, $g\to U_g$, with $U_e=\Id$, is a projective representation of $G$ if the following conditions hold.
\begin{description}
\item[\;{\rm i)}\;\;]
A complex function $\sigma:{G}\times{G}\to{\CC}$,
called multiplier,
exists such that $U_{g_1g_2}=\sigma(g_1,g_2)U_{g_1}U_{g_2}$;
\item[\;{\rm ii)}\;]
for all $\phi,\psi\in\mathcal H$, the mapping $g\to\langle U_g\phi\mid\psi\rangle$ is a Borel function in $g$.
\end{description}\noindent
A projective representation with multiplier $\sigma$ is also called $\sigma$-representation.
\par
A projective representation is said to be continuous if for any fixed $\psi\in\mathcal H$
the mapping $g\to U_g\psi$ from $G$ to $\mathcal H$ is continuous with respect to $g$.}
\vskip.5pc\noindent
Let ${\mathcal E}$ be the Euclidean group, i.e. the
semi-direct product ${\mathcal E}={\RR}^3\sm SO(3)$ between the group of spatial translations ${\RR}^3$ and
the group of spatial proper rotations $SO(3)$;
each transformation $g\in\mathcal E$ bi-univocally corresponds to the pair $({\bf a},R)\in {\RR}^3\times SO(3)$
such that $R^{-1}{\bf x}-R^{-1}{\bf a}\equiv \textsf g({\bf x})$ is the result of the passive
transformation of the spatial point ${\bf x}=(x_1,x_2,x_3)$ by $g$.
The general imprimitivity theorem is an advanced mathematical result \cite{c4};
in this article we shall make use of this theorem relatively to the euclidean group $\mathcal E$ only.
Then we introduce the concept of {\sl imprimitivity system} and the theorem for this specific case \cite{c3},\cite{c5}.
\vskip.5pc\noindent
{\bf Definition 2.2.} {\sl
Let ${\mathcal H}$ be the Hilbert space of a $\sigma$-representation
$g\to U_g$ of the Euclidean group $\mathcal E$. A {\sl projection valued} (PV) measure
$E:{\mathcal B}({\RR}^3)\to \Pi({\mathcal H})$, $\Delta\to E(\Delta)$
is an {\sl imprimitivity system} for the $\sigma$-representation $g\to U_g$ if the
relation
$$U_g E(\Delta)U_g^{-1}=E(\hbox{\rm $\textsf g^{-1}$}(\Delta))\equiv E(R(\Delta)+{\bf a})\eqno(1)$$
holds for all $({\bf a},R)\in\mathcal E$.}
\vskip.5pc\noindent
{\bf Mackey's theorem of imprimitivity for $\mathcal E$.}
{\sl
If a PV measure $E:{\mathcal B}({\RR}^3)\to \Pi({\mathcal H})$
is an imprimitivity system for a continuous $\sigma$-representation $g\to U_g$ of the Euclidean group
$\mathcal E$, then a $\sigma$-representation $L:SO(3)\to{\mathcal U}({\mathcal H}_0)$ exists
such that, modulo a unitary isomorphism,
\vskip.3pc\noindent
(M.1) ${\mathcal H}=L_2({\RR}^3,{\mathcal H}_0)$, \vskip.3pc\noindent
(M.2) $(E(\Delta)\psi)({\bf x})=\chi_{_\Delta}
({\bf x})\psi({\bf x})$, where $\chi_{_\Delta}$ is the characteristic functional of $\Delta$,
\vskip.3pc\noindent
(M.3) $(U_g\psi)({\bf x})=L_R\psi(\hbox{\rm\textsf g}({\bf x}))\equiv L_R\psi(R^{-1}{\bf x}-R^{-1}{\bf a})$,
for every $g=({\bf a},R)\in\mathcal E$.
\vskip.3pc\noindent
Furthermore, the $\sigma$-representation $U$ is irreducible if and only if the
``inducing'' representation $L$ is irreducible.}\vskip.5pc
\subsection{Basic concepts}
In this subsection we formulate a concept of quantum transformation, which is viable also for space-time transformations that
are not symmetry transformations.
\par
For sake of synthesis, in the following by $\Upsilon$ we denote the group $\mathcal G$ of galileian transformations
without time translations and space-time inversions,
or the group $\mathcal P$ of Poincar\'e's transformations without space-time inversions;
therefore, what stated for $\Upsilon$ must be understood stated
for $\mathcal G$ and also for $\mathcal P$. In the present work
the group $\Upsilon$ is interpreted as a group of changes of
reference frame in a class $\mathcal F$ of frames which move uniformly with respect to each other.
So, given any reference frame $\Sigma$ in $\mathcal F$, a transformation $g\in\Upsilon$ univocally singles out the reference frame
$\Sigma_g$ related to $\Sigma$ just by $g$.
\par
Let us consider the Quantum Theory of a {\sl localizable particle}, that is to say of a physical system which
can be localized in a point of the physical space, so that its Quantum Theory contains
a unique triple $(Q_1,Q_2,Q_3)\equiv\bf Q$ of
commuting self-adjoint operators  representing the three coordinates of the position. Now,
the point of the space, where the particle is localized by the measurement of the position observables,
is identified only if the frame is specified the values of the coordinates refer to.
For instance,
if $(Q_1,Q_2,Q_3)\equiv\bf Q$ are the three self-adjoint
operators which represent the three coordinates
of the position with respect to $\Sigma$ and if $g\in\mathcal E$, then the $\alpha$-th coordinate of the position
with respect to another frame $\Sigma_g$, related to $\Sigma$ by $g$, must be represented by
$[\textsf g({\bf Q})]_\alpha$, where $\textsf g({\bf x})=(y_1,y_2,y_3)$ is the
triple of the coordinates, with respect to $\Sigma_g$, of the spatial point represented by $\bf x$ with respect to $\Sigma$.
In the non relativistic case, a pure galileian boost $g\in\mathcal G$ characterized by a velocity ${\bf u}=(u,0,0)$,
does not change the instantaneous position at all; hence \rm{$\textsf g({\bf x})={\bf x}$ and
\rm{$S^\Sigma_g[{\bf Q}]=\textsf g({\bf Q})={\bf Q}$, so that the operators which represent the coordinates of the
``position with respect to $\Sigma_g$'' coincide with the operators representing the position coordinates
with respect to $\Sigma$.
In order to transform the
position quantum observables at time $t$}, i.e. the operators ${\bf Q}^{(t)}=e^{iHt}{\bf Q}^{(t)}e^{-iHt}$,
by a galileian boost $g$, a function {\rm{${\textsf g}_t$}} different from {\rm{$\textsf g$}} must be used.
Indeed, ${\bf Q}^{(t)}$ represents the position measured with a delay $t$, therefore
the operators which represent the ``position at time $t$ with respect to $\Sigma_g$''
must be
\rm{$S_g^\Sigma[{\bf Q}^{(t)}]=(Q_1^{(t)}-ut,Q_2,Q_3)\equiv\textsf g_t({\bf Q}^{(t)})$}, where
\rm{$\textsf g_t({\bf x})=(x_1-ut,x_2,x_3)$}\par
In general, we can state that for every $g\in\mathcal G$ the following covariance relations hold for all $g\in\mathcal G$,
$$
(i)\quad
S^\Sigma_g[{\bf Q}]=\hbox{\rm{\textsf g}}({\bf Q}),
\qquad(ii)\quad S^\Sigma_g[{\bf Q}^{(t)}]=\hbox{\rm{\textsf g}}_t({\bf Q}^{(t)})
,\eqno(2)
$$
where {\rm{\textsf g}}$_t$ is a suitable function, in general different from {\rm{\textsf g}}.
In fact, relations (2) are the conditions which {\sl define} the position operators of a localizable particle.
\vskip.5pc
{\it A priori} we cannot exclude that also observables other than position change their representation
according to the frame they are referred to; so,
in order that the Quantum Theory of our particle can account for such a possibility, it must appropriately extend
the transformations $S_g^\Sigma$ to all quantum observables.
To this aim, given two reference frames $\Sigma_1$ and $\Sigma_2$ in $\mathcal F$, we introduce the following concept of
{\sl relative indistinguishability} between measuring procedures:
\vskip.4pc
{\sl If a measuring procedure ${\mathcal M}_1$ is relatively to $\Sigma_1$ identical to what is ${\mathcal M}_2$ relatively to
$\Sigma_2$, we say that ${\mathcal M}_1$ and ${\mathcal M}_2$ are indistinguishable relatively to $(\Sigma_1,\Sigma_2)$.}
\vskip.4pc\noindent
Then, for every $g\in\Upsilon$ and every $\Sigma$ in $\mathcal F$ we introduce the mapping
$$
S_g^\Sigma:\Omega({\mathcal H})\to \Omega({\mathcal H}),\quad A\to S_g^\Sigma[A]\eqno(3)
$$
with the following conceptually explicit interpretation.
\begin{description}
\item[{\rm
(QT)}]{\sl\;
The self-adjoint operators $A$ and $S_g^\Sigma[A]$ represent two measuring procedures
${\mathcal M}_1$ and ${\mathcal M}_2$ indistinguishable relatively to $(\Sigma,\Sigma_g)$.
}
\end{description}
For instance, if $A$ represents a detector placed in the origin of $\Sigma$ with a given orientation relative to $\Sigma$,
then $S_g^\Sigma[A]$ is the operator
that represents an identical detector placed in the origin of $\Sigma_g$ with that orientation relative to $\Sigma_g$.
It must be noticed that (QT) presupposes that for each quantum observable $A\in\Omega({\mathcal H})$ and every $g\in\Upsilon$, two measuring
procedures with the required relative indistinguishability exist, at least in principle.
\par
We call $S^\Sigma_g$ the {\sl quantum transformation}
corresponding to $g$.
\vskip.5pc\noindent
Relations (2) explicitly specify the action of the transformations $S_g^\Sigma$ on the position
operators ${\bf Q}^{(t)}$; for an arbitrary observable no such a kind of explicit specification can be {\it a priori}
established.
However, the authentic meaning (QT) of the notion of quantum transformation is sufficient to infer,
at a conceptual level, the following general constraint.
\begin{description}
\item[{\rm(S.1)}]
{\sl For every frame $\Sigma$ in $\mathcal F$ the following statement holds.
$$
S_{gh}^\Sigma[A]=S_g^{\Sigma_h}\left[S_h^\Sigma[A]\right],\hbox{ for all }A\in\Omega(\mathcal H).\eqno(4)
$$}
\end{description}
This statement stresses how in general, i.e. without further particular conditions, the
mapping $S_g^\Sigma$, with $g$ fixed, can change by changing the ``starting'' frame $\Sigma$.
\subsection{Symmetry transformations}
Let us now briefly outline the particular stronger implications of the existence of conditions of symmetry.
A transformation $h\in\Upsilon$ is a symmetry transformation for the physical system under investigation if
a class $\mathcal F$ exists such that
for every frame $\Sigma$ in $\mathcal F$, the frames $\Sigma$ and $\Sigma_h$ are
equivalent for the formulation of the empirical theory of the system; for an isolated system, all $g\in\Upsilon$
are symmetry transformations.
\vskip.5pc
The symmetry property allows to apply Wigner's theorem, and in so doing
the following well known implication is obtained
\cite{b1},\cite{b2},\cite{c9},\cite{c10}.
\vskip.5pc\noindent
\textsc{Sym.1}.
{\sl
If $g\in\Upsilon$ is a symmetry transformation then a unitary or an anti-unitary operator
$U_g^\Sigma$, unique up a phase factor, exists such that
$$S_g^\Sigma[A]=U_g^\Sigma A\left[U_g^\Sigma\right]^\ast.\eqno(5)$$}
Moreover, according to the Principle of Relativity, for an isolated system all $g\in\Upsilon$ are symmetry transformations.
Therefore,
a class $\mathcal F$ exists such that the following statement holds.
\vskip.5pc\noindent
\textsc{Sym.2}.
{\sl
In the Quantum Theory of an isolated system, for each $g\in\Upsilon$ the quantum transformation $S_g^\Sigma$ must be independent of
\,$\Sigma$, i.e. $S_g^\Sigma=S_g^{\Sigma_h}\equiv S_g$ and $U^\Sigma_g=e^{i\lambda}U^{\Sigma_h}_g$ (with $\lambda\in\RR$),
so that (4) and (5) imply
$$
S_{gh}[A]=S_g\left[S_h[A]\right]\,.\eqno(6)
$$}
Therefore, $U_{gh}=\sigma(g,h)U_gU_h$ holds, which implies that each $U_g$ is unitary \cite{b5},\cite{c9};
in particular, $U_g^\ast=U_g^{-1}$.
Thus, if $\Upsilon$ is a group of symmetry transformations, the correspondence
$g\to U_g$ such that $S_g[A]=U_gAU_g^\ast$ is a projective representation
\cite{b4},\cite{b5},\cite{c11}.
\vskip.5pc
A {\it free localizable particle} is just a particular kind of isolated system, so that
according to \textsc{Sym.2} for every $g\in\Upsilon$
a unitary operator $U_g$ exists such that $S_g[A]=U_gA U_g^{-1}$.
The restriction of $g\to U_g$ to the euclidean group $\mathcal E$  is a projective
representation of $\mathcal E$ \cite{b5}. Then,
according to (2) and \textsc{Sym.1}, the relation $U_g{\bf Q}U_g^{-1}=${\rm \textsf g}$({\bf Q})$
holds;
it entails that the common spectral PV spectral measure of ${\bf Q}=(Q_1,Q_2,Q_3)$ is an imprimitivity system
for $U\mid_{\mathcal E}$ \cite{b5}; therefore we can apply Mackey's imprimitivity theorem. In so doing,
to each choice of the inducing representation $L:SO(3)\to{\mathcal U}({\mathcal H}_0)$
in Mackey's theorem and of $\mu$ in (9),
there corresponds a different theory.
Accordingly, the Hilbert space of the theory can be identified as
$L_2(\RR^3,{\mathcal H}_0)$, and the position operators as
$(Q_\alpha\psi)({\bf x})=x_\alpha\psi({\bf x})$. Furthermore,
in a non-relativistic theory, by making use of
Galileian invariance, valid for a free particle, it can be proved \cite{b5},\cite{c12} that the
form of the hamiltonian operator must be $H=-\frac{1}{2\mu}\sum_{\alpha=1}^3\frac{\partial^2}{\partial x_\alpha^2}$.
By choosing $L$ as an irreducible $\sigma$-representation of $SO(3)$ of dimension $2s+1$ ($s\in\frac{1}{2}\NN$),
the Standard Quantum Theory of a spin-$s$ free particle is obtained.
\subsection{The interacting particle problem}
If the system under investigation is not isolated, e.g. if it is an {\sl interacting} particle,
then neither \textsc{Sym.1} nor \textsc{Sym.2} apply, so that
we find an obstacle in extending the group theoretical approach to the non-relativistic interacting particle.
However, in the literature several proposals can be found \cite{b5},\cite{c7},\cite{c8},\cite{c6}
where the group theoretical methods are extended to the interacting case. Each proposal overcomes
the aforesaid obstacles, in the non-relativistic case, by introducing particular assumptions we reformulate into the following
statement.
\vskip.5pc\noindent
\textsc{Proj}.\,
{\sl
Each galileian transformation $g$ is represented in the formalism of the Quantum Theory
by a unitary operator $U_g$ in such a way that}
\item[\;\;{\rm i)}]{\sl
$S_g^\Sigma[{\bf Q}^{(t)}]=U_g{\bf Q}^{(t)}U_g^{-1}$ is the quantum transformation of the
``position at time $t$'' observables corresponding to $g$;
\item[\;{\rm ii)}]
the correspondence $g\to U_g$ is a continuous projective representation.}
\vskip.5pc\noindent
Statement \textsc{Proj} is introduced as an assumption in \cite{b5}, page 201; the conditions assumed in \cite{c7} by Jauch, page 236,
are implied by \textsc{Proj};
Ekstein, instead, essentially derives it from another assumption,
namely from the
``empirical statement that it is possible to give an operational definition of any initial state
intrinsically'', i.e. independently of the presence or absence of the interaction (cfr. \cite{c6}, page 1401).
\par
By making use of
\textsc{Proj}, some of the cited approaches \cite{b5},\cite{c7} deduce that
in the non-relativistic Quantum Theory of a spin-0 particle, undergoing an interaction homogeneous in time,
the hamiltonian operator $H$ must have the following form, able to describe also interactions of electromagnetic kind
\cite{b5},\cite{c7}.
$$H=\frac{1}{2\mu}\sum_{\alpha=1}^3\left(-i\frac{\partial}{\partial x_\alpha}+
a_\alpha({\bf x)}\right)^2+\Phi({\bf x}).\eqno(7)$$
Now we shall prove,
instead, the following statement.
\vskip.5pc\noindent
\textsc{Stat}.\,
{\sl
Assumption
\textsc{Proj} implies that the hamiltonian of the Quantum Theory of a spin-0
particle undergoing an interaction homogeneous in time must have the form}
$$
H=\frac{1}{2\mu}\sum_{\alpha=1}^3\left(-i\frac{\partial}{\partial x_\alpha}\right)^2+\Phi({\bf x}).$$
To prove the sentence $\textsc{Stat}$ we shall make use of the following well known
results.
\vskip.5pc\noindent
{\bf MP.1.}\,
As an important implication of Wigner's theorem, the general
evolution law of quantum observables with respect to a homogeneous time is obtained \cite{c9}:
a self-adjoint operator $H$ exists, called hamiltonian operator, such that
$$
A^{(t)}=e^{iHt}Ae^{-iHt}\quad\hbox{and}\quad \frac{d}{dt}A^{(t)}\equiv\dot A^{(t)}=i[H,A^{(t)}].\eqno(8)
$$
{\bf MP.2.}\,
Let $g\to\hat U_g$ be
every continuous non trivial projective representation
of Galilei's group $\mathcal G$, i.e. the group generated by the euclidean
group $\mathcal E$ and by galileian velocity boosts.
Now, the nine one-parameter abelian sub-groups ${\mathcal T}_\alpha,{\mathcal R}_\alpha$,
${\mathcal B}_\alpha$ of spatial translation, spatial rotations
and galileian velocity boosts, relative to axis $x_\alpha$, are all additive; then,
according to Stone's theorem \cite{c9}, there exist
nine self-adjoint generators $\hat P_\alpha$, $\hat J_\alpha$, $\hat G_\alpha$ of the nine
one-parameter unitary subgroups $\{e^{-i{\hat P}_\alpha a_\alpha},\,a\in{\RR}\}$,
$\{e^{-i{\hat J}_\alpha \theta_\alpha},\,\theta_\alpha\in{\RR}\}$, $\{e^{-i{\hat G}_\alpha u_\alpha},\,u_\alpha\in{\RR}\}$
that represent the
sub-groups ${\mathcal T}_\alpha,{\mathcal R}_\alpha$, ${\mathcal B}_\alpha$ according to the projective representation
$g\to {\hat U}_g$ of the Galilei's group $\mathcal G$.
The
structural properties of $\mathcal G$ as a Lie group imply the following
commutation relations \cite{c13}.
\vskip.4pc\noindent
(i) $[{\hat P}_\alpha,{\hat P}_\beta]=\nop$,\quad (ii) $[{\hat J}_\alpha,{\hat P}_\beta]=i{\hat\epsilon}_{\alpha\beta\gamma}{\hat P}_\gamma$,
\quad (iii) $[{\hat J}_\alpha,{\hat J}_\beta]=i{\hat\epsilon}_{\alpha\beta\gamma}{\hat J}_\gamma$,\par\noindent
(iv) $[{\hat J}_\alpha,{\hat G}_\beta]
=i{\hat\epsilon}_{\alpha\beta\gamma}{\hat G}_\gamma$,\quad (v) $[{\hat G}_\alpha,{\hat G}_\beta]=\nop$,
\quad (vi) $[{\hat G}_\alpha,{\hat P}_\beta]=i\delta_{\alpha\beta}\mu\Id$,\hfill{}(9)
\vskip.4pc\noindent
where ${\hat\epsilon}_{\alpha,\beta,\gamma}$ is the Levi-Civita symbol ${\epsilon}_{\alpha,\beta,\gamma}$
restricted by the condition $\alpha\neq\gamma\neq\beta$, and
$\mu$ is a non-zero real number which characterizes the projective representation.
\vskip.5pc\noindent
\textsc{Proof of Stat.}\;
Now we explicitly prove \textsc{Stat}. Since
$g\to U_g$ in \textsc{Proj} is a projective representation, according to ({\bf MP.2})
the sub-groups ${\mathcal T}_\alpha$, ${\mathcal R}_\alpha$, ${\mathcal B}_\alpha$
can be represented by the one-parameter unitary sub-groups
$\{e^{-iP_\alpha a}\}_{a\in{\RR}}$, $\{e^{-iJ_\alpha\theta}\}_{\theta\in{\RR}}$,
$\{e^{iG_\alpha u}\}_{u\in{\RR}}$, in such a way that the self-adjoint generators
$P_\alpha,J_\alpha,G_\alpha$ satisfy (9).
Once defined the self-adjoint operators $F_\alpha=\frac{G_\alpha}{\mu}$,
it can be proved that relations (9) imply that
the following relation holds for all $g\in\mathcal G$.
$$U_g{\bf F}U_g^{-1}=\textsf g({\bf F}).\eqno(10)$$
Since by (9.v) the $F_\alpha$'s commute with each other, according to spectral theory, a unique
PV measure
$E:{\mathcal B}({\RR}^3)\to\Pi({\mathcal H})$ exists
such that $F_\alpha=\int\lambda dE^{(\alpha)}_\lambda$, where $E^{(1)}_\lambda=E((-\infty,\lambda]\times{\RR}^2)$,
$E^{(2)}_\lambda=E({\RR}\times(-\infty,\lambda]\times{\RR})$, $E^{(3)}_\lambda=E({\RR}^2\times(-\infty,\lambda])$.
Then (10) easily implies that $\Delta\to E(\Delta)$ satisfies (1) and hence
it is an {\sl imprimitivity} system for the {\sl restriction} to
$\mathcal E$ of $g\to U_g$; therefore Mackey's theorem applies.
In so doing, the simplest choice for ${\mathcal H}_0$, i.e. ${\mathcal H}_0={\CC}$,
leads to identify $\mathcal H$, $F_\alpha$, $P_\alpha$, and $U_g$ for $g\in\mathcal E$ as
$$
{\mathcal H}=L_2({\RR}^3),\quad (F_\alpha\psi)({\bf x})=x_\alpha\psi({\bf x}),\quad
P_\alpha=-i\frac{\partial}{\partial x_\alpha},\quad
(U_g\psi)({\bf x})=\psi\left(\textsf g({\bf x})\right).\eqno(11)$$
Now
we can easily prove
that the position operators $\bf Q$ coincide with ${\bf F}={\bf G}/\mu$.
\vskip.5pc\noindent
{\bf Proposition 2.1.}
{\sl If \textsc{Proj} holds, then in the simplest Quantum Theory of a localizable interacting
particle the equality ${\bf F}={\bf Q}$ holds for the position operators satisfying the covariance properties (2).}
\vskip.3pc\noindent
{\bf Proof.}
If $g\in{\mathcal T}_\beta$ and \textsc{Proj} holds, so that by ({\bf MP.2})
$U_g=e^{-iP_\beta a}$, then
(2.i) implies $[Q_\alpha,P_\beta]=i\delta_{\alpha\beta}\Id$;
since $[F_\alpha,P_\beta]=i\delta_{\alpha,\beta}\Id$ is implied by (9.iv), also
$[F_\alpha-Q_\alpha,P_\beta]=\nop$ holds.
On the other hand (2.i) for $U_g=e^{iG_\beta u}$ implies
$[F_\alpha-Q_\alpha,F_\beta]=\nop$, and hence $F_\alpha-Q_\alpha=c_\alpha\Id\equiv$constant
must hold for the irreducibility of $({\bf F, P})$. Finally, \textsc{Proj}.i
together with (9.iv) and (2.i) for
$U_g=e^{-iJ_\alpha\theta}$ imply
$[J_\alpha,F_\beta-Q_\beta]=i{\hat\epsilon}_{\alpha,\beta,\gamma}(F_\gamma-Q_\gamma)=i{\hat\epsilon}_{\alpha,\beta,\gamma}c_\gamma\Id=[J_\alpha,c_\beta\Id]=\nop$; thus,
$F_\alpha-Q_\alpha=\nop$.{\hfill{$\bullet$}}
\vskip.5pc
Prop. 2.1
together with (2.ii) is sufficient to determine the form of the hamiltonian
operator $H$ consistent with {\textsc{Proj}}.
First, we determine $[G_\alpha,\dot Q_\beta]$.
Let us start with
$$
e^{iG_\alpha u}\dot Q_\beta e^{-iG_\alpha u}=\dot Q_\beta+i[G_\alpha,\dot Q_\beta]u+o(u),\eqno(12)
$$
where $o(u)$ is an infinitesimal operator of order greater than 1 with respect to $u$.
By making use of $\dot Q_\beta=i[H,Q_\beta]=\lim_{t\to 0}\frac{Q_\beta^{(t)}-Q_\beta}{t}$, and of
$e^{iG_\alpha u}Q_\beta^{(t)}e^{-iG_\alpha u}=Q_\beta^{(t)}-\delta_{\alpha\beta}ut\Id$, implied by
(2.ii),
we also find
$$e^{iG_\alpha u}\dot Q_\beta e^{-iG_\alpha u}=\lim_{t\to 0}e^{iG_\alpha u}
\frac{Q_\beta^{(t)}-Q_\beta}{t}e^{-iG_\alpha u}=
\dot Q_\beta-\delta_{\alpha\beta}u\Id.\eqno(13)
$$
The comparison between (12) and (13), and Prop. 2.1 yield
$$
[G_\alpha,\dot Q_\beta]=[Q_\alpha,\mu\dot Q_\beta]=i\delta_{\alpha\beta}\Id,\hbox{ which implies }
[F_\alpha,\mu\dot Q_\beta-P_\beta]=\nop.\eqno(14).
$$
This argument can be repeated with $U_g=e^{-iP_\alpha a}$ instead of $e^{iG_\alpha u}$, and also with
$U_g=e^{-iJ_\alpha\theta}$ instead of $e^{iG_\alpha u}$. In so doing we obtain, respectively,
$[P_\alpha,\mu\dot Q_\beta-P_\beta]=\nop$ and $[J_\alpha,\mu\dot Q_\beta]=i{\hat\epsilon}_{\alpha,\beta,\gamma}\mu\dot Q_\gamma$;
the first of these two equations,
together with (14), implies $\mu\dot Q_\beta-P_\beta=b_\beta\Id$; then, by making use of the second
equation, we obtain
$i\hat\epsilon_{\alpha\beta\gamma}(\mu\dot Q_\gamma-P_\gamma)=[J_\alpha,\mu\dot Q_\beta-P_\beta]=[J_\alpha,b_\beta\Id]=\nop$,
i.e.
$\mu\dot Q_\beta=P_\beta$.
\vskip.5pc
At this point the determination of $H$ is straightforward. From (9.vi) we obtain
$$
i[H,Q_\beta]=\dot Q_\beta=\frac{1}{\mu}P_\beta=i\left[\frac{1}{2\mu}\sum_\gamma P_\gamma^2,\frac{G_\beta}{\mu}\right]
\equiv i\left[\frac{1}{2\mu}\sum_\gamma P_\gamma^2,Q_\beta\right]\,
.\eqno(15)
$$
Then the completeness of $\bf Q$ implies that the operator $H-\frac{1}{2\mu}\sum_\gamma P_\gamma^2$ is a function $\Phi$ of $\bf Q$.
Thus
$$
H=-\frac{1}{2\mu}\left(\frac{\partial^2}{\partial x_1^2}+\frac{\partial^2}{\partial x_2^2}+\frac{\partial^2}{\partial x_3^2}\right)+\Phi({\bf Q}).
\eqno(16)
$$
Thus,
assumption \textsc{Proj} forbids the description of electromagnetic interactions,
because their physics is correctly described by the hamiltonian in
(7), inequivalent to (16).
\section{Quantum Theory of an interacting particle}
Coherently with the conclusion of the last section,
in order to develop a Quantum Theory of a particle
able to describe also electromagnetic interactions,
assumption \textsc{Proj}
must be abandoned.
In this section we undertake such a development, under the hypothesis that the interaction does not destroy time homogeneity,
so that according to ({\bf MP.1}) an hamiltonian operator $H$ exists such that (8) holds
\vskip.5pc\noindent
We begin by identifying two further properties (S.2) and (S.3) of quantum transformations, which add to
the general property (S.1) already established.
\begin{description}
\item[{\rm(S.2)}]
For every $g\in\Upsilon$, the mapping $S^\Sigma_g$ is bijective.
\item[{\rm(S.3)}]
For every real Borel function $f$ such that if $A$ is a self-adjoint operator,
then $B=f(A)$ is a self-adjoint operator too, the following equality holds:
$$
f(S^\Sigma_g[A])=S^\Sigma_g[f(A)].
\eqno(17)
$$
\end{description}
In fact,
these further properties are directly
implied on a conceptual ground by the  meaning of quantum transformation expressed by (QT).
For instance, with regard to (S.3), one can argue as follows.
Let $f$ be any fixed real Borel function such that if $A$ is a self-adjoint operator,
then $B=f(A)$ is a self-adjoint operator too.
Now, according to Quantum Theory a measurement of the quantum observable $f(A)$
can be performed by measuring $A$ and then
transforming the obtained outcome $a$ by the purely mathematical function
$f$ into the outcome $b=f(a)$ of $f(A)$.
If a measurement procedure is relatively to $\Sigma$ identical to another measuring procedure relatively to $\Sigma_g$, then
transforming the outcomes of both procedures by means of the same function $f$ should not affect the relative indistinguishability of the
so modified procedures. So we should conclude that (17) holds.\par
Hence, the concept (QT)  entails the  validity of (S.2) and (S.3);
for reasons we shall indicate later in remark 3.1, however,
for the time being we formulate them as conditions which characterize a class of interactions.
\vskip.5pc\noindent
In sections 3.1, 3.2 we show that the further properties (S.2), (S.3) imply that if the correspondence
$g\to S_g^\Sigma$ is continuous according to the metric adopted by Bargmann \cite{c11},
then for every $g\in\Upsilon$ a {\sl unitary} operator $U_g$ must exists such that
\par
i) $\;g\to U_g$ is continuous;
\par
ii) $S_g^\Sigma[A]=U_gAU_g^{-1}$.
\par\noindent
This result addresses one of the obstacles against the extension of the
group theoretical approach to an interacting particle;
but other obstacles remain. Indeed, in order to explicitly identify the
mathematical formalism of the theory we should apply the imprimitivity theorem;
but this is not possible because, while the mapping $g\to U_g$ is continuous under a condition of continuity for
$g\to S_g^\Sigma$, it is not a projective representation, and such a projectivity condition is required by the imprimitivity theorem.
\par
To address this new obstacle we shall introduce in section 3.3 the notion of ``$\sigma$-conversion'', which
is a consistent mathematical process carrying out the conversion of the mapping
$U:\Upsilon\to {\mathcal U}({\mathcal H})$, $g\to U_g$ into a mapping
$\hat U:\Upsilon\to\hat{\mathcal U}({\mathcal H})$, $g\to \hat U_g$ which is a projectiove representation.
\par
Using $\sigma$-conversions shall allow the approach to proceed. In the non-relativistic case, where $\Upsilon=\mathcal G$,
we prove that the postion operators $\bf Q$ coincide with the multiplication operators endowed with the usual
interpretation if and only if the interaction admits
``Q-covariant'' $\sigma$-conversions, i.e. $\sigma$-conversions that leave unaltered
the covariance properties of the position operators $\bf Q$ with respect to $\mathcal G$.
For Q-covariant $\sigma$-conversions we derive a general dynamical equation (27), in section 3.4.
\subsection{General implications for quantum transformations}
Conditions (S.2) and (S.3), are sufficient to show further properties
of the mappings $S^\Sigma_g$, according to the following Prop.s 3.1 and 3.2.
\vskip.5pc\noindent
{\bf Proposition 3.1.}
{\sl
Let $S:\Omega({\mathcal H})\to\Omega({\mathcal H})$ be a bijective mapping
such that $S[f(A)]=f(S[A])$ for every Borel real function $f$ such that $f(A)\in\Omega({\mathcal H})$ if $A\in\Omega({\mathcal H})$.
Then the following statements hold.
\begin{description}
\item[\;\;{\rm i)}]
If $E\in\Pi({\mathcal H})$ then $S[E]\in\Pi[{\mathcal H}]$, i.e., the mapping $S$ is an extension
of a bijection of $\Pi({\mathcal H}]$.
\item[\;{\rm ii)}]
If $A,B\in\Omega({\mathcal H})$ and $A+B\in\Omega({\mathcal H})$, then $[A,B]=\nop$ implies
$S[A+B]=S[A]+S[B]$.
\item[\;\;\;]
This partial additivity immediately implies $S[A]=\nop$ if and only if $A=\nop$.
\item[{\rm iii)}]
For all
$E,F\in\Pi({\mathcal H})$, $EF=\nop$ implies $S[E+F]=S[E]+S[F]\in\Pi({\mathcal H})$;
as a consequence,
$E\leq F$ if and only if $S[E]\leq S[F]$.
\item[{\rm iv)}]
$S[P]\in\Pi_1({\mathcal H})$ if and only if $P\in\Pi_1({\mathcal H})$.
\end{description}
}\noindent
{\bf Proof.}\quad(i)
If $E\in\Pi({\mathcal H})$ and $f(\lambda)=\lambda^2$ then $f(E)=E$ holds; so
$S[f(E)]=f(S[E])$ implies $(S_g[E])^2\equiv f(S[E])=S[E^2]\equiv S[E]$, i.e. $S^2[E]=S[E]$.
\par
(ii)
If $[A,B]=\nop$ then a self-adjoint operator $C$ and two functions $f_a$, $f_b$ exist
so that $A=f_a(C)$ and $B=f_b(C)$; once defined the function $f=f_a+f_b$, we have
$S[A+B]\equiv S[f(C)]=f(S[C])=f_a(S[C])+f_b(S[C])=S[f_a(C)]+S[f_b(C)]\equiv S[A]+S[C]$.
\par
(iii)
If $EF=\nop$, then $[E,F]=\nop$ and $(E+F)\in\Pi({\mathcal H})$ hold.
Statements (i) and (ii) imply $S[E+F]=S[E]+S[F]\in\Pi({\mathcal H}]$.
\par
(iv)
If $P\in\Pi_1({\mathcal H})$ then $S[P]\in\Pi({\mathcal H})$ by (i). If $Q\in\Pi_1({\mathcal H})$ and $Q\leq S[P]$
then $P_0\equiv S^{-1}[Q]\leq P$ by (iii); but $P$ is rank 1, therefore $P_0=P$ and $Q=S[P]$.
\hfill{$\bullet$}
\vskip.5pc\noindent
{\bf Corollary 3.1.} {\sl From Prop.3.1 immediately follows that the
restriction of $S$ to $\Pi({\mathcal H})$ is a bijection that also satisfies
$S[\nop]=\nop$, $S[\Id ]=\Id $, $E\leq F$ iff $S[E]\leq S[F]$,
$S[E^\bot]=(S[E])^\bot$.}
\vskip.5pc\noindent
In the literature different equivalent formulations of Wigner's theorem
\cite{b2},\cite{c14} have been proved. The following version shall find application for the the mapping $S$ of Prop. 3.1.
\vskip.8pc\noindent
{\bf Wigner's theorem.}
{\sl If $S:\Pi({\mathcal H})\to\Pi({\mathcal H})$ is an automorphism of $\Pi({\mathcal H})$, i.e. if it is
a bijective mapping such that
$$E_1\leq E_2\Leftrightarrow S[E_1]\leq S[E_2]\quad\hbox{and}\quad S[E^\bot]=(S[E])^\bot,\quad
\forall E_1,E_2,E\in\Pi({\mathcal H}),$$
then either a unitary operator or an anti-unitary operator $U$ of $\mathcal H$ exists such that $S(E)=UEU^{\ast}$ for all $E\in\Pi({\mathcal H})$,
unique up a phase factor.}
\vskip.5pc\noindent
In virtue of Corollary 3.1 and of Wigner's theorem, the following proposition is easily proved.
\vskip.4pc\noindent
{\bf Proposition 3.2.}
{\sl
If a mapping $S$ satisfies the hypothesis of Prop. 3.1, then
a unitary or an anti-unitary operator exists such that $S[A]=UAU^\ast$ for every $A\in\Omega({\mathcal H})$;
if another unitary or anti-unitary operator $V$ satisfies $S[A]=VAV^\ast$ for every $A\in\Omega(\mathcal H)$, then
$V=e^{i\theta}U$ with $\theta\in{\RR}$.}
\vskip.5pc
Prop.s 3.1 and 3.2 are proved for a mapping $S:\Omega({\mathcal H})\to\Omega({\mathcal H})$; therefore they hold
for every quantum transformation $S_g^\Sigma$ of the Quantum Theory of a particle whose interaction is in the class for which (S.2), (S.3) hold. Then,
for each $g\in\Upsilon$, according to Prop. 3.2
each transformation $g\in\Upsilon$ is assigned a unitary or an anti unitary operator
$U_g$ which realizes the corresponding quantum transformation as the automorphism
$S_g^\Sigma:\Pi({\mathcal H})\to\Pi({\mathcal H})$,
$S^\Sigma_g[A]=U_gAU_g^\ast$, also if $g$ is not a symmetry transformation.
\subsection{Continuity and unitarity of $g\to U_g$}
Given $g\in\Upsilon$, the unitary or anti-unitary operator $U_g$ such that $S^\Sigma_g[A]=U_gAU_g^\ast$
can be arbitrarily chosen within an equivalence class ${\bf U}_g$ of operators, all unitary or all anti-unitary,
which differ from each other by a complex phase factor; this class ${\bf U}_g$ is called {\sl
operator ray} \cite{c11}; due to Wigner's theorem, there is a bijective correspondence between operator rays and
automorphisms of $\Pi({\mathcal H})$. The possibility that the choice of $U_g$ within ${\bf U}_g$ makes the correspondence
$g\to U_g$ continuous has a decisive role in developing the Quantum Theory of a physical system; for instance, for the
non-relativistic Quantum Theory of a free particle, it makes possible Stone's theorem to apply, and as a
consequence the one-parameter sub-groups ${\mathcal T}_\alpha$, ${\mathcal R}_\alpha$, ${\mathcal B}_\alpha$
can be represented as $e^{-i P_\alpha a}$, $e^{-i J_\alpha\theta}$, $e^{i G_\alpha u}$.
According to results due to Bargmann \cite{c11}, a choice of $U_g$ in ${\bf U}_g$ leading to a continuous correspondence $g\to U_g$ exists
if
the mapping $g\to S^\Sigma_g$ is {\sl continuous}, where $S_g^\Sigma:\Pi({\mathcal H})\to\Pi({\mathcal H})$ is the
restriction to $\Pi({\mathcal H})$ of the quantum transformation corresponding to $g$.
However, Bargmann carried out his proof by requiring that {\sl all operators $U_g$ are unitary}. Now we see how the implication
proved by Bargmann holds also if such a restriction is removed.
\par
The continuity notion of Bargmann\footnote{In fact Bargmann's continuity refers to
a correspondence $g\to{\bf U}_g$ from a topological group $G$ to the set of all {\sl unitary}
operator rays ${\bf U}_g$; but, since an operator ray can be bijectively identified with
an automorphism of $\Pi({\mathcal H})$, Bargmann's continuity can be reformulated in terms of automorphisms;
this reformulation immediately extends to all automorphisms, included those corresponding to anti-unitary
operator rays, through our Def. 3.2.}
for $g\to S_g^\Sigma$
is based on the following metric of $\Pi_1({\mathcal H})$.
\vskip.5pc\noindent
{\bf Definition 3.1.}
{\sl
Given two rank 1 projection operators $D_1,D_2\in\Pi_1({\mathcal H})$, the distance $d(D_1,D_2)$ is the minimal distance $\Vert\psi_1-\psi_2\Vert$
between vectors $\psi_1,\psi_2$ such that $P_1=\vert\psi_1\rangle\langle\psi_1\vert$ and
$P_2=\vert\psi_2\rangle\langle\psi_2\vert$, i.e.,
$d(D_1,D_2)=[2(1-\vert\langle\psi_1\mid\psi_2\rangle\vert]^{1/2}$.}
\vskip.5pc\noindent
Then, following Bargmann, the continuity of a mapping from a topological group $G$ to the automorphisms of $\Pi({\mathcal H})$,
is defined as follows.
\vskip.5pc\noindent
{\bf Definition 3.2.}
{\sl A correspondence $g\to S_g$ from a topological group $G$ to the set of all automorphisms of $\Pi({\mathcal H})$
is said to be continuous if for any fixed $D\in\Pi_1({\mathcal H})$ the mapping
from $G$ to $\Pi_1({\mathcal H})$, $g\to S_g[D]$ is continuous in $g$ with respect to
the distance $d$ defined on $\Pi_1({\mathcal H})$ by Def. 3.1.}
\vskip.4pc\noindent
Before proving the main result Prop.3.3, we formulate three lemmas. The first one, Lemma 3.1,
was proved by Bargmann as Lemma 1.1 in \cite{c11}.
\vskip.5pc\noindent
{\bf Lemma 3.1.}
{\sl
The real function $\kappa:\Pi_1({\mathcal H})\times\Pi_1({\mathcal H})\to\RR$, $\kappa(D_1,D_2)=Tr(D_1D_2)$
is continuous in both variables $D_1$ and $D_2$ with respect to the metric of Def. 3.1.}\vskip.5pc\noindent
{\bf Lemma 3.2.}
{\sl Given a topological group $G$ and a mapping $g\to S_g$ from $G$ to the automorphisms of $\Pi({\mathcal H})$,
for every $g\in G$ let ${\bf U}_g$ denote the operator ray identified by $S_g$;
for every $\varphi\in\mathcal H$ with $\Vert\varphi\Vert=1$, let us define
$$
z_{h,g}(\varphi)=U_g\varphi-\langle U_h\varphi\mid U_g\varphi\rangle U_h\varphi,
$$
where $h,g\in G$, $U_h\in{\bf U}_h$ and $U_g\in{\bf U}_g$. Then
$$
\Vert z_{h,g}(\varphi)\Vert^2=1-\vert\langle U_h\varphi\mid U_g\varphi\rangle\vert^2\leq d^2(S_h[D_\varphi],S_g[D_\varphi]);
$$
where $D_\varphi=\vert\varphi\rangle\langle\varphi\vert\in\Pi_1({\mathcal H})$.}
\vskip.4pc\noindent
{\bf Proof.}
The proof is identical to the proof of statement (1.9) in Theorem 1.1 of \cite{c11};
indeed that proof can be successfully carried out independently of the unitary or anti-unitary character of $U_g$ or $U_{h}$.
\vskip.5pc\noindent
{\bf Lemma 3.3.}
{\sl Let $G$ be a topological group, let $g\to S_g$ be a {\it continuous} mapping from $G$ to the automorphisms of
$\Pi({\mathcal H})$, and let us fix an operator $U_g\in{\bf U}_g$ for each $g\in G$.
\par
If $U_g\varphi_0$ is continuous in $g$ as a function from $G$ to $\mathcal H$ for
a vector $\varphi_0\in\mathcal H$ with $\Vert\varphi_0\Vert=1$,
then $U_g\varphi_1$ is continuous in $g$ for every fixed $\varphi_1\in\mathcal H$ with $\Vert\varphi_1\Vert=1$, such that $\varphi_1\perp\varphi_0$.}
\vskip.4pc\noindent
{\bf Proof.}
We prove the lemma by adapting a part of the proof of Theorem 1.1
in \cite{c11}. Let us define
$\varphi=\frac{1}{\sqrt{2}}(\varphi_0+\varphi_1)$; of course we have
$\langle U_g\varphi_0\mid U_g\varphi\rangle=\frac{1}{\sqrt{2}}$ for all $g\in G$ independently of the unitary
or anti-unitary character of $U_g\in{\bf U}_g$. Then
\begin{eqnarray*}
\langle U_h\varphi_0\mid z_{h,g}(\varphi)\rangle
&=&\langle U_h\varphi_0-U_g\varphi_0\mid U_g\varphi\rangle+\langle U_g\varphi_0\mid U_g\varphi\rangle
-\langle U_h\varphi\mid U_g\varphi\rangle\langle U_h\varphi_0\mid U_h\varphi\rangle\\
&=&\langle U_h\varphi_0-U_g\varphi_0\mid U_g\varphi\rangle+\frac{1}{\sqrt{2}}(1-\langle U_h\varphi\mid U_g\varphi\rangle.
\end{eqnarray*}
So
$$
(1-\langle U_h\varphi\mid U_g\varphi\rangle)=
\sqrt{2}\left\{\langle U_h\varphi_0\mid z_{h,g}(\varphi)\rangle+\langle U_g\varphi_0-U_h\varphi_0\mid U_g\varphi\rangle\right\}\;.\eqno(18)
$$
Now,
\begin{eqnarray*}
\Vert U_g\varphi-U_h\varphi\Vert^2
&=& 2\vert \RRe(1-\langle U_h\varphi\mid U_g\varphi\rangle)\vert\leq 2\vert 1-\langle U_h\varphi\mid U_g\varphi\rangle\vert\\
&\leq&
2\sqrt{2}\{\vert\langle U_h\varphi_0\mid z_{h,g}(\varphi)\rangle\vert+2\sqrt{2}
\vert U_g\varphi_0-U_h\varphi_0\mid U_g\varphi\rangle\vert\}\\
&\leq&
2\sqrt{2}\Vert\mid z_{h,g}(\varphi)\Vert+2\sqrt{2}
\Vert U_g\varphi_0-U_h\varphi_0\Vert\\
&\leq& 2\sqrt{2}\left( d(S_h[D_\varphi],S_g[D_\varphi])+\Vert U_g\varphi_0-U_h\varphi_0\Vert\right)\;,
\end{eqnarray*}
where we made use of (18) in the second inequality, in the third inequality we used
Schwarz inequality, and in the fourth inequality Lemma 3.2 is applied.
These inequalities imply that $U_g\varphi$ is continuous in $g$; indeed,
the distance $d(S_h[D_\varphi],S_g[D_\varphi])$ vanishes as $g\to h$ because the mapping $g\to S_g$ is continuous according to
Def. 3.2 by the first continuity hypothesis; but also $\Vert U_g\varphi_0-U_h\varphi_0\Vert$ vanishes as $g\to h$, because $U_g\varphi_0$ is continuous in $g$
by the second continuity hypothesis.
\par
Now, $\varphi_1=\sqrt{2}\varphi-\varphi_0$, so that $U_g\varphi_1 =\sqrt{2}U_g\varphi_2-U_g\varphi_0$ for all $g$ such that $U_g$
is unitary, but also for all $g$ such that $U_g$ is anti-unitary. Thus $U_g\varphi_1$ is continuous because
$U_g\varphi$ and $U_g\varphi_1$ are continuous.\hfill{$\bullet$}
\vskip.5pc\noindent
Let us arbitrarily fix a vector $\varphi_0\in\mathcal H$, with $\Vert\varphi_0\Vert=1$. Given any mapping
$g\to S_g$ from a topological group $G$ to the automorphisms of $\Pi({\mathcal H})$, we define
the real function $\rho_{\varphi_0}:G\to\RR$, $\rho_{\varphi_0}(g)={Tr}^{1/2}(D_{\varphi_0}S_g[D_{\varphi_0}])$.
Since $S_g[D_{\varphi_0}]={\tilde U}_gD_{\varphi_0}{\tilde U}_g^\ast$, where ${\tilde U}_g$ is any operator in ${\bf U}_g$, we have
$\rho_{\varphi_0}(g)=\vert\langle\varphi_0\mid{\tilde U}_g\varphi_0\rangle\vert$. Hence,
$\langle\varphi_0\mid{\tilde U}_g\varphi_0\rangle=e^{i\alpha(g)}\vert\langle\varphi_0\mid{\tilde U}_g\varphi_0\rangle\vert
=e^{i\alpha(g)}\rho_{\varphi_0}(g)$, for some $\alpha(g)\in\RR$. Then
$\rho_{\varphi_0}(g)=\vert\langle\varphi_0\mid{\tilde U}_g\varphi_0\rangle\vert=e^{-i\alpha(g)}\langle\varphi_0\mid{\tilde U}_g\varphi_0\rangle$.
Therefore, if for each $g\in G$ we choose $U_g=e^{-i\alpha(g)}{\tilde U}_g$ we obtain
$$
\rho_{\varphi_0}(g)=\langle\varphi_0\mid{U}_g\varphi_0\rangle\;;\hbox{ in particular, }\ U_e=\Id\;.\eqno(19)
$$
{\bf Proposition 3.3.}
{\sl Let $G$ be a topological group, and let $\varphi_0$ be any fixed vector in $\mathcal H$ with $\Vert\varphi_0\Vert=1$. Given
a mapping $g\to S_g$ from $G$ to the automorphisms of $\Pi({\mathcal H})$, if each $g\in G$ is assigned
the operator $U_g\in{\bf U}_g$ such that (19) holds, then $U_g\psi$ is continuous in $g$, whatever be the vector $\psi\in\mathcal H$.}
\vskip.4pc\noindent
{\bf Proof.}
Bargmann proved that if $g\to S_g$ is continuous according to Def. 3.2 and if $U_g$ is the operator such that (19) holds, then $U_g\varphi_0$
is continuous\footnote{In fact, Bargmann proved this statement for unitary $U_g$; but Bargmann's proof can be successfully
carried out without assuming that all $U_g$ are unitary.}.
Now, let $\psi$ be any vector of $\mathcal H$.
\par
If $\psi=0$,
then the continuity of $U_g\psi$ is obvious. Therefore it is sufficient to prove the proposition for $\psi\neq 0$.
\par
If $\psi=\lambda\varphi_0$ for some $\lambda\in\CC\setminus\{0\}$, then we choose any $\varphi_1\perp\varphi_0$,
with $\Vert\varphi_1\Vert=1$. According to Lemma 3.3, $U_g\varphi_1$ is continuous. The same Lemma implies
that $U_g\frac{\psi}{\Vert\psi\vert}$ is continuous because $\frac{\psi}{\Vert\psi\Vert}\perp\varphi_1$. But
$U_g\psi=\Vert\psi\Vert U_g\frac{\psi}{\Vert\psi\Vert}$ for all $g\in G$. Therefore $U_g\psi$ is continuous.
\par
If $\psi\neq\lambda\varphi_0$, define $\varphi=\frac{\psi}{\Vert\psi\Vert}$; then a vector $\varphi_1\in\mathcal H$ exists, with $\Vert\varphi_1\Vert=1$ and
$\varphi_1\perp\varphi_0$, such that
$$
\varphi=a\varphi_0+b\varphi_1\quad\hbox{where } a\in\CC\hbox{ but }b\in\RR\,.\eqno(20)
$$
Now, a real number
$r$ and a vector $\varphi_2$, with $\Vert\varphi_2\Vert=1$ exist such that $a\varphi_0=r\varphi_2$; this implies $\varphi_2\perp\varphi_1$
and $\varphi=r\varphi_2+b\varphi_1$.
Lemma 3.3 implies that $U_g\varphi_1$ is continuous because $\varphi_1\perp\varphi_0$; but the same Lemma implies that also
$U_g\varphi_2$ is continuous, because $\varphi_2\perp\varphi_1$. Therefore, since $r$ and $b$ are real numbers,
$U_g\varphi=r U_g\varphi_2+b U_g\varphi_1$ is continuous in $g$. Thus, $U_g\psi=\Vert\psi\Vert U_g\varphi$ is continuous too.
\hfill{$\bullet$}
\vskip.8pc
Another condition with helpful implications is the unitary character of the operators $U_g$ that realize the quantum transformations
according to $S_g^\Sigma[A]=U_gAU_g^{-1}$. If the correspondence $g\to S_g^\Sigma$ satisfied
$S_{g_1g_2}^\Sigma=S_{g_1}^\Sigma\circ S_{g_2}^\Sigma$ so that
$g\to U_g$ would be a projective representation, then
it could be easily proved, according to \cite{b3},\cite{b5},\cite{c9},\cite{c11}, that
every $U_g$ must be unitary. But in presence of interaction $S_g^{\Sigma_1}$ can be different from $S_g^{\Sigma_2}$, so that
only the more general statement (S.1) holds, and hence the unitary character of
$U_g$ cannot be implied by the cited proofs.
Now we prove that anti-unitary $U_g$ can be excluded under the only hypothesis
that the correspondence $g\to S_g^\Sigma$ is continuous according to Def. 3.2.
\vskip.5pc\noindent
{\bf Proposition 3.4.}
{\sl If the mapping $g\to S^\Sigma_g$, that assigns each $g\in\Upsilon$ the quantum transformation of (3), is continuous
according to Def. 3.2, then for every operator $U_g$ such that $S^\Sigma_g[A]=U_gAU_g^\ast$ for all $A\in\Omega({\mathcal H})$
is unitary.}
\vskip.4pc\noindent
{\bf Proof.}
According to Prop. 3.3, for every $g\in\Upsilon$ a unitary or anti-unitary operator such that
$S^\Sigma_g[A]=U_gAU_g^\ast$ exists which makes $U_g\psi$ continuous in $g$ for all $\psi$.
According to (19) $U_e=\Id$ which is unitary. Hence, because of the continuity
of $g\to U_g\psi$ for all $\psi$, a maximal neighborhood $K_e$ of $e$ must exist in $\Upsilon$ such that $U_g$ is unitary for all $g\in K_e$;
otherwise a sequence $g_n\to e$ would exist with $U_{g_n}$ anti-unitary, so that
$\langle\psi\mid\varphi\rangle=\langle U_{g_n}\varphi \mid U_{g_n}\psi\rangle$ for all $\psi,\varphi\in\mathcal H$, and then
$\langle\psi\mid\varphi\rangle=\lim_{n\to\infty}\langle U_{g_n}\varphi \mid U_{g_n}\psi\rangle
=\langle U_e\varphi\mid U_e\psi\rangle=\langle\varphi\mid\psi\rangle$. This last equality cannot hold for all $\psi,\varphi\in\mathcal H$
unless ${\mathcal H}$ is real.
\par
Now we prove that such a neighborhood $K_e$ has no boundary, and since $\Upsilon$ is a connected group,
$K_e=\Upsilon$. If
$g_0\in\partial K_e$, two sequences $g_n\to g_0$ and $h_n\to g_0$ would
exist with $U_{g_n}$ unitary and $U_{h_n}$ anti-unitary; therefore, the continuity of $U_g$
would imply that $U_{g_0}$ should simultaneously be unitary and anti-unitary. \hfill{$\bullet$}
\vskip.5pc\noindent
{\bf Remark 3.1.}
{\sl
The work of this subsection has shown that (S.2) and (S.3) imply that $S_g^\Sigma[A]=U_gAU_g^{-1}$,
where $U_g$ is unitary if $g\to S_g^\Sigma$ is continuous; as a consequence, the spectrum of any quantum observable is left
unchanged by $S_g^\Sigma$.
\par
Such an invariance has important consequences; for instance,
it entails that particular kinds of interactions are not compatible with the theory.
Indeed,
let the interaction be able to confine a localizable particle in a bounded region of the physical space.
For sake of simplicity, we assume that space is one dimensional, so that there is only one position operator $Q$ whose possible
values are confined by the interaction in the interval $[0,a]$ of the only axis of the reference frame $\Sigma$;
this means that $[0,a]$ contains the spectrum of $Q$, of course: $\sigma(Q)\subseteq [0,a]$.
Let $g\in\Upsilon$ be
the spatial translation identified by ${\textsf g}(x)=x-a$.
According to
(2) we have $S_g^\Sigma[Q]=Q-a$, and hence $\sigma(S_g^\Sigma[Q])\subseteq [-a,0]$:
$S_g^\Sigma$ changes the spectrum of $Q$.}
\vskip.5pc\noindent
Therefore, if (S.2) and (S.3) were generally valid constraints,
the confinement interaction should not be an interaction compatible with Quantum Theory,
i.e. no interaction could sharply confine a particle within a bounded region.
Such a drastic conclusion is based on (S.2) and (S.3) which, though endowed with conceptual soundness,
do not have a formal derivation. For this reason we find appropriate, for the time being,
to establish (S.2) and (S.3) as conditions which characterize the class of interactions investigated in the present
work.
In the following we shall see that such a class is a very large one, enough, in particular, to encompass electromagnetic interaction.}
\subsection{$\sigma$-conversions}
In section 3.2 we have established, under a continuity condition for $g\to S^\Sigma_g$,
that in the Quantum Theory of a physical system, also if it is not isolated, a continuous
correspondence
$U:\Upsilon \to\mathcal U(\mathcal H)$ exists such that $S^\Sigma_g[A]=U_gAU_g^{-1}$.
To assume that such a correspondence is a projective representation implies \textsc{Proj}; therefore, according to section 2.4,
the out-coming theory is unable to describe particles interacting with electromagnetic fields.
So, we give up this condition with the scope of developing a Quantum Theory of an interacting particle empirically more adequate.
But without such a ``projectivity'' condition Mackey's imprimitivity theorem does not apply. Hence,
the development of our group-theoretical approach
encounters a further obstacle. Now we address this obstacle.
\par
The correspondence $g\to U_g$,
can be {\sl converted} into a continuous $\sigma$-representation
if we multiply each operator $U_g$ by a suitable unitary operator $V_g$
of $\mathcal H$; namely, $V_g$ is a unitary operator such that
the correspondence $g\to \hat U_g=V_gU_g$ turns out to be a $\sigma$-representation.
The transition from the correspondence $\{g\to U_g\}$ to $\{g\to\hat U_g=V_gU_g\}$ will be called
{\sl $\sigma$-conversion}; the mapping $V:\Upsilon\to {\mathcal U}({\mathcal H})$,
$g\to V_g$  that realizes the $\sigma$-conversion will be called $\sigma$-{\sl conversion mapping}.
If $g\to V_g$ is a $\sigma$-conversion mapping for $g\to U_g$ and
$\theta:\Upsilon\to {\RR}$ is a real function, then also $g\to e^{i\theta(g)}V_g$
is a $\sigma$-conversion mapping, provided that $e^{i\theta(e)}=1$.
In any case, $V_e=\Id $ must hold.
\par
Since non-trivial projective representations of $\Upsilon$ exist,we can state the following proposition.
\vskip.5pc\noindent
{\bf Proposition 3.5.}
{\sl
A correspondence $V:\Upsilon\to{\mathcal U}({\mathcal H})$ always exists such that
$\hat U:\Upsilon\to{\mathcal U}({\mathcal H})$, $g\to\hat U_g=V_gU_g$ is
a non trivial projective representation.}
\vskip.5pc
The $\sigma$-conversion allows to immediately identify a mathematical formalism for
the Quantum Theory of the system, also in the case that the system is not isolated.
In the case of a non-relativistic theory, where $\Upsilon=\mathcal G$,
if $g\to V_g$ is a $\sigma$-conversion mapping for $U_g$ then, according to
({\bf MP.2}) in sect. 2.4, the $\sigma$-representation
$g\to\hat U_g=V_gU_g$ has nine hermitean
generators $\hat P_\alpha$, $\hat J_\alpha$, $\hat G_\alpha$ for which (9) hold.
Then, following the argument of the proof of \textsc{Stat} in section 2.4,
the common spectral measure of the triple ${\bf F}=\hat{\bf G}/\mu$
turns out to be an imprimitivity system for the restriction of
$g\to\hat U_g$ to $\mathcal E$.
So, by applying the imprimitivity theorem of Mackey \cite{b5}, we can explicitly identify $\mathcal H$
as $L_2(\RR^3,{\mathcal H}_0)$, modulo unitary isomorphisms, where the operators
$F_\alpha$, $\hat P_\alpha$, $\hat J_\alpha$ and $\hat G_\alpha$ are explicitly specified according to
$${\mathcal H}=L_2({\RR}^3,{\mathcal H}_0)\,,\qquad
(F_\alpha\psi)({\bf x})=x_\alpha\psi({\bf x})\,,\qquad
\hat P_\alpha=-i\frac{\partial}{\partial x_\alpha}\,,\eqno(21)$$
$$
\hat J_\alpha=F_\beta\hat P_\gamma-F_\gamma\hat P_\beta+S_\alpha\,,\qquad
\hat G_\alpha=\mu F_\alpha\,.
$$
Here $(\alpha,\beta,\gamma)$ is a cyclic permutation of (1,2,3);
the $S_\alpha$ are operators that act on ${\mathcal H}_0$ only, i.e. their action is
$(S_\alpha\psi)({\bf x})=\hat s_\alpha\psi({\bf x})$ where the $\hat s_\alpha$
are self-adjoint operators of ${\mathcal H}_0$
which form a representation of the commutation rules
$[\hat s_\alpha,\hat s_\beta]=i{\hat\epsilon}_{\alpha\beta\gamma}\hat s_\gamma$.
Since the reducibility of the inducing representation $L:SO(3)\to{\mathcal U}({\mathcal H}_0)$ implies the
reducibility of $\hat U:{\mathcal G}\to{\mathcal U}({\mathcal H})$,
if $\hat U$ is irreducible then also $(\hat s_1,\hat s_2,\hat s_3)$ must be an irreducible representation
of $[\hat s_\alpha,\hat s_\beta]=i{\hat\epsilon}_{\alpha\beta\gamma}\hat s_\gamma$; in this case,
modulo unitary isomorphisms, ${\mathcal H}_0$ is one of the
finite-dimensional Hilbert spaces $\,\CC^{2s+1}$, with $s\in\frac{1}{2}\NN$: the $\hat s_\alpha$ are the familiar spin operators.
\vskip.5pc
Hence, the mathematical formalism of the Quantum Theory of a localizable particle
has been explicitly identified.
However, the operators $\hat U_g$ concretely identified
{\sl are not} the unitary operators which realize the
quantum transformations: given $g\in\mathcal G$, in general $S^\Sigma_g[A]=\hat U_gA\hat U_g^{-1}$ does not hold.
As a consequence the operators ${\bf Q}=(Q_1,Q_2,Q_3)$ representing the position
cannot be identified following section 2.3 or the argument of the proof of Prop. 2.1.
So, our explicit realization of the mathematical formalism of
the theory would be, in general, devoid of physical significance.
\par
Two tasks have to be accomplished in order that the formalism established by (21)
becomes the mathematical formalism
of the effective Quantum Theory of an interacting particle.
\par
First, the operators $\bf Q$ of the Hilbert space ${\mathcal H}=L_2(\RR^3,{\mathcal H}_0)$ in (21),
that represent the position of the particle, should be explicitly determined. We address this task in section 3.4.
\par
Second, the wave equation ruling over the time evolution should be determined.
In section 3.5 we derive a general dynamical law. Specific wave equations corresponding to
specific features of the interaction are determined in section 4.
\subsection{Q-covariant $\sigma$-conversions}
The position operators $\bf Q$ can be determined for those interactions that have the particular feature
of admitting a $\sigma$-conversion $U_g\to \hat U_g=V_g U_g$
that {\sl leaves unaltered the covariance properties of the position operators $\bf Q$}, i.e. such that
$$
\hat U_g{\bf Q}\hat U_g^{-1}= \textsf g({\bf Q})\,,\;\forall g\in{\mathcal G}.\eqno(22)
$$
A $\sigma$-conversion satisfying (22) is said to be {\sl Q-covariant}.
Indeed, the following proposition hold.
\vskip.4pc\noindent
{\bf Proposition 3.6.}\,
{\sl
If a $\sigma$-conversion for a particle yields an irreducible projective
representation $\hat U$,
then it is a Q-covariant $\sigma$-conversion if and only if
the position operators $\bf Q$ coincide with $\bf F$.}
\vskip.4pc\noindent
{\bf Proof.}
If ${\bf Q}={\bf F}=\hat{\bf G}/\mu$, then (9) imply
$\hat U_g{\bf Q}\hat U_g^{-1}\equiv\hat U_g{\bf F}\hat U_g^{-1}=\textsf g({\bf F})=\textsf g({\bf Q})$.\par
Conversely,
if $\hat U:\mathcal G\to\mathcal U(\mathcal H)$ is an irreducible projective
representation obtained from $U:\mathcal G\to\mathcal U(\mathcal H)$
through a Q-covariant $\sigma$-conversion,
then (22) for $\hat U_g=e^{i\hat G_\beta u}=e^{i\mu F_\beta u}$ and (9.v) imply
$[Q_\alpha-F_\alpha,F_\beta]=[Q_\alpha,F_\beta]-[F_\alpha,F_\beta]=\nop-\nop=\nop$; therefore
$(Q_\alpha-F_\alpha)\psi({\bf x})=\left(f_\alpha({\bf Q})\psi\right)({\bf x})=f_\alpha({\bf x})\psi({\bf x})$,
where $f_\alpha({\bf x})$ is a self-adjoint operator of
$\mathcal H_0$. However, the Q-covariance and (9.vi) imply also
$[Q_\alpha-F_\alpha,\hat P_\beta]=[Q_\alpha,\hat P_\beta]-[Q_\alpha,\hat P_\beta]=
i\delta_{\alpha\beta}\Id-i\delta_{\alpha\beta}\Id=\bf 0$,
i.e. $[f_\alpha({\bf Q}),\hat P_\beta]=\bf 0$ for all $\bf x$;
this relation, since $\hat P=-i\frac{\partial}{\partial x_\alpha}$, implies that
$\frac{\partial f_\alpha}{\partial x_\alpha}({\bf x})=0$, for all $\alpha,\beta$;
therefore $f_\alpha({\bf x})$ is an operator $\hat f_\alpha$ of
$\mathcal H_0$ which {\sl does not depend} on $\bf x$.
Now, since $\hat f_\alpha =Q_\alpha-F_\alpha$, also $[\hat f_\alpha,\hat f_\beta]=\bf 0$ holds;
moreover, from (2.i) for a pure spatial rotation $g$ about $x_\alpha$ and from (9.iv) we obtain
$[\hat J_\alpha,Q_\beta-F_\beta]=i\hat\epsilon_{\alpha\beta\gamma}(Q_\gamma-F_\gamma)=i\hat\epsilon_{\alpha\beta\gamma} \hat f_\gamma$;
but the irreducibility of $\hat U$ implies the irreducibility of the inducing projective representation
$L:SO(3)\to{\mathcal U}({\mathcal H}_0)$, so that
$\mathcal H_0$ is finite dimensional; then
$[\hat f_\alpha,\hat f_\beta]=\bf 0$ and $[\hat J_\alpha,\hat f_\beta]=i\hat\epsilon_{\alpha\beta\gamma} \hat f_\gamma$
can hold only if $\hat f_\alpha=\bf 0$, i.e. $F_\alpha=Q_\alpha$.
\hfill{$\bullet$}
\vskip.5pc
Hence, in the Quantum Theory of an interacting particle, where $\hat U$ is irreducible,
the multiplication operators can be identified with the position operators
if and only if the interaction has the particular regularity feature of admitting a
$\sigma$-conversion which preserves the covariance properties of the position operators.
\par
Following a customary habit, we say that a particle, whose interaction admits Q-covariant $\sigma$-conversion,
is {\sl elementary} if $\hat U$ is irreducible.
\vskip.5pc
The following proposition specify how in the Quantum Theory of an elementary particle
each $\hat U_g$ is related to the unitary operator $U_g$ that realizes
the quantum transformation corresponding to $g$.
\vskip.4pc\noindent
{\bf Proposition 3.7.}\,
{\sl For every $g\in\mathcal G$, the operator $V_g$ of a Q-covariant $\sigma$-conversion has the form
$(V_g\psi)({\bf x})=\left(e^{i\theta(g,{\bf Q})}\psi\right)({\bf x})=e^{i\theta(g,{\bf x})}\psi({\bf x})$,
where $\theta(g,{\bf x})$ is a self-adjoint operator of $\mathcal H_0$ which depends on $\bf x$ and on $g$.}
\vskip.4pc\noindent
{\bf Proof.}
Relations(22) and (2) imply
$V_gU_g{\bf Q}U_g^{-1}V_g^{-1}=\textsf g({\bf Q})$, which implies
$V_g (\textsf g({\bf Q}))V_g^{-1}=\textsf g({\bf Q})$, i.e. $[V_g ,\textsf g({\bf Q})]=\nop$. Then
$[V_g,\textsf f( \textsf g({\bf Q}))]=\nop$ for every sufficiently regular function $\textsf f$;
by taking $\textsf f=\textsf g^{-1}$ we have
$[V_g,{\bf Q}]=\bf 0$. Then $(V_g\psi)({\bf x})=\textsf h_g({\bf x})\psi({\bf x})$, where
$\textsf h_g({\bf x})$ is an operator of $\mathcal H_0$.
Finally, the unitary character of $V_g$ imposes that $h_g({\bf x})$ must be unitary as an operator
of ${\mathcal H}_0$; thus
a self-adjoint operator $\theta(g,{\bf x})$ of ${\mathcal H}_0$
exists such that $\textsf h_g({\bf x})=e^{\theta(g,{\bf x})}$.\hfill{$\bullet$}
\vskip.5pc
If $g\to S_g^\Sigma$ is continuous according to Def. 3.2, then
$g\to V_g$ must be continuous because $g\to{\hat U}_g=V_gU_g$ is continuous.
\vskip.5pc\noindent
{\bf Remark 3.2.}
{\sl
In the present approach the imprimitivity system for applying Mackey's theorem
is identified within the abstract projective representation itself, namely it is the PV spectral measure of
$\hat{\bf G}/\mu$.
This is a remarkable difference with respect to the past approaches, e.g. Mackey's approach,
where the imprimitivity system is identified as the PV measure of the position operators.}
\subsection{General Dynamical law}
Now we derive a general
dynamical equation ruling over the time evolution of an elementary particle.
In so doing we shall suppose that the $\sigma$-conversion mapping $g\to V_g$ is differentiable with respect to
the parameters $a_\alpha,\theta_\alpha,u_\alpha$ of the group $\mathcal G$.
\par
Let us consider the pure velocity boost $g\in\mathcal G$ such that $\hat U_g=e^{i\hat G_\alpha u}$.
According to sect. 3.3, the formalism of its Quantum Theory
can be identified with that established by (21).
Since $\hat G_\alpha=\mu F_\alpha=\mu Q_\alpha$, we can write $\hat U_g=e^{i\mu Q_\alpha u}$; therefore
$$
\hat U_g\dot Q_\beta \hat U_g^{-1}=\dot Q_\beta+i\mu[Q_\alpha,\dot Q_\beta]u+o_1(u).\eqno(23)
$$
On the other hand,
$$
\hat U_g\dot Q_\beta \hat U_g^{-1}=\lim_{t\to 0}V_gU_g\frac{(Q_\beta^{(t)}-Q_\beta)}{t}U_g^{-1}V_g^{-1}.\eqno(24)
$$
By making use of
$U_gQ_\beta^{(t)}U_g^{-1}= Q_\beta^{(t)}-\delta_{\alpha\beta}ut\Id$,
implied by (2), and of Prop. 3.2, Prop. 3.3 in (24), and then
comparing with (23) we obtain
$$
\hat U_g\dot Q_\beta \hat U_g^{-1}=V_g\dot Q_\beta V_g^{-1}-\delta_{\alpha\beta}u\Id=
\dot Q_\beta+i\mu[Q_\alpha,\dot Q_\beta]u+o_1(u).\eqno(25)
$$
But Prop. 3.7 implies that $V_g=e^{i\varsigma_\alpha(u,{\bf Q})}$,
where $\varsigma_\alpha(u,{\bf x})$ is a self-adjoint operator of $\mathcal H_0$;
replacing in (25) we obtain
$$
\dot Q_\beta +i[\varsigma_\alpha(u,{\bf Q}),\dot Q_\beta]+o_2(u)-\delta_{\alpha\beta}u\Id
=\dot Q_\beta+i\mu[Q_\alpha,\dot Q_\beta]u+o_1(u).\eqno(26)
$$
Since $e^{i\varsigma_\alpha(0,{\bf Q})}=\Id$,
the expansion  of $\varsigma_\alpha$ with respect to $u$ yields
$\varsigma_\alpha(u,{\bf Q})=\frac{\partial\varsigma_\alpha}{\partial u}(0,{\bf Q})u+o_3(u)$; by replacing this
last relation in (26) we obtain
$$
\mu[Q_\alpha,\dot Q_\beta]=[\eta_\alpha({\bf Q}),\dot Q_\beta]+i\delta_{\alpha\beta}\Id,
$$
where $\eta_\alpha({\bf Q})=\frac{\partial\varsigma_\alpha}{\partial u}(0,{\bf Q})$.
By replacing
$\dot Q_\beta=i[H,Q_\beta]$
in this last equation we can apply Jacobi's identity, and in so doing we obtain
$[Q_\beta,\mu\dot Q_\alpha]=[Q_\beta,\dot\eta_\alpha({\bf Q})]+i\delta_{\alpha\beta}\Id$,
i.e.
$$[Q_\beta,\dot\eta_\alpha({\bf Q})-\mu\dot Q_\alpha]=-i\delta_{\alpha\beta}\Id=[Q_\beta,-\hat P_\alpha].$$
Hence $[\dot\eta_\alpha({\bf Q})-\mu\dot Q_\alpha-\hat P_\alpha,Q_\beta]=\bf 0$, from which
we imply that for every ${\bf x}\in\RR^3$ an operator $f_\alpha({\bf x})$ of $\mathcal H_0$, must exist
such that the equation
$\{\dot\eta({\bf Q})-\mu \dot Q_\alpha+\hat P_\alpha\}\psi({\bf x})=f_\alpha({\bf x})\psi({\bf x})$
holds, that we can rewrite as
$$
i[H,\mu Q_\alpha-\eta_\alpha({\bf Q})]=\hat P_\alpha-f_\alpha({\bf Q}).\eqno(27)
$$
This is a general dynamical equation for a localizable particle whose interaction admits Q-covariant $\sigma$-conversions;
according to such a law, the effects of the interaction on the dynamics are encoded in the six ``fields'' $\eta_\alpha$, $f_\alpha$.
\subsection{Electromagnetic interaction for spin-0 particles}
Once derived the general dynamical law (27) for an elementary particle
with homogeneous in time interaction, it is worth
to re-discover the wave equation currently adopted in quantum physics as
a particular case of the general equation (27).
In this subsection we do this for
a spin-0 particle, for which ${\mathcal H}_0={\CC}$ so that ${\mathcal H}=L_2({\RR}^3)$.
The nowadays adopted Schroedinger equation for a spin-0 particle has the form
$$
i\frac{d}{dt}\psi_t=\left\{\frac{1}{2 m}
\sum_{\alpha=1}^3[\hat P_\alpha+a_\alpha({\bf Q})]^2+\Phi({\bf Q})\right\}\psi_t,\eqno(28)
$$
i.e. the Hamiltonian operator is $H=(1/2\mu)\sum_{\alpha=1}^3\{\hat P_\alpha+a_\alpha({\bf Q})\}^2+\Phi({\bf Q})$,
where $a_\alpha({\bf Q})$ and $\Phi({\bf Q})$ are self-adjoint operators of $L_2(\RR^3)$ functions of $\bf Q$.
Now we show that within our approach this specific Quantum Theory bi-univocally
corresponds to the case that the functions $\eta_\alpha$ in the general law (27) are constant functions multiple of $\Id$.
\vskip.5pc\noindent
{\bf Proposition 3.8.}
{\sl
The hamiltonian operator $H$ of an interacting spin-0 particle which admits $Q$-covariant
$\sigma$-conversion has the form
$H=(1/2\mu)\sum_{\alpha=1}^3\{\hat P_\alpha+a_\alpha({\bf Q})\}^2+\Phi({\bf Q})$
if and only if
the functions $\eta_\alpha$ in (27) are constant functions.
In this case $a_\alpha=-f_\alpha$.}
\vskip.4pc\noindent
{\bf Proof.}
If $\eta_\alpha$ is a constant function, then (27) transforms  into $i[H,\mu Q_\alpha]=\hat P_\alpha-f_\alpha({\bf Q})$ which holds
if $H_0=\frac{1}{2\mu}\sum_{\alpha=1}^3\{\hat P_\alpha-f_\alpha({\bf Q})\}^2$ replaces $H$. Hence the operator $H-H_0$
must be a function $\Phi$ of $\bf Q$ because of the completeness of $\bf Q$. Then $\eta_\alpha({\bf Q})=c_\alpha\Id$ implies
$H=\frac{1}{2\mu}\sum_{\alpha=1}^3[\hat P_\alpha-f_\alpha({\bf Q})]^2+\Phi({\bf Q})$.
\par
Now we prove the converse.
Let us suppose that
$H=\frac{1}{2\mu}\sum_{\alpha=1}^3\{\hat P_\alpha+a_\alpha({\bf Q})\}^2+\Phi({\bf Q})$;
by replacing this $H$ in (27) we obtain\vskip.6pc\noindent
$i[H,\mu Q_\alpha-\eta_\alpha({\bf Q})]=\hat P_\alpha-f_\alpha({\bf Q})=$
\begin{eqnarray*}
&=&\frac{i}{2\mu}\sum_\beta[\hat P^2_\beta,\mu Q_\alpha]+\frac{i}{2\mu}\sum_\beta[a_\beta\hat P_\beta,\mu Q_\alpha]+
\frac{i}{2\mu}\sum_\beta[\hat P_\beta a_\beta,\mu Q_\alpha]+\frac{i}{2\mu}\sum_\beta[a^2_\beta,\mu Q_\alpha]+\\
&&-
\frac{i}{2\mu}\sum_\beta[\hat P^2_\beta,\eta_\alpha]-\frac{i}{2\mu}\sum_\beta[a_\beta\hat P_\beta,\eta_\alpha]-
\frac{i}{2\mu}\sum_\beta[\hat P_\beta a_\beta,\eta_\alpha]-\frac{i}{2\mu}\sum_\beta[a^2_\beta,\eta_\alpha]+\\
&&+i[\Phi({\bf Q}),\mu Q_\alpha-\eta_\alpha].
\end{eqnarray*}
In the last member of these equalities, the fourth, the eighth and the
last term are zero. Then we have
\vskip.5pc\noindent
$i[H,\mu Q_\alpha-\eta_\alpha({\bf Q})]=\hat P_\alpha-f_\alpha({\bf Q})$\hfill{(29)}
\begin{eqnarray*}
&=&\hat P_\alpha+\frac{i}{2}\sum_\beta
(a_\beta\hat P_\beta Q_\alpha-Q_\alpha a_\beta \hat P_\beta+\hat P_\beta a_\beta Q_\alpha-Q_\alpha\hat P_\beta a_\beta)+\\
&&-\frac{i}{2\mu}\sum_\beta[\hat P^2_\beta,\eta_\alpha]-
\frac{i}{2\mu}\sum_\beta
(a_\beta\hat P_\beta \eta_\alpha-\eta_\alpha a_\beta \hat P_\beta+\hat P_\beta a_\beta \eta_\alpha-\eta_\alpha\hat P_\beta a_\beta)\\
&=&\hat P_\alpha+\frac{i}{2}\sum_\beta
(a_\beta[\hat P_\beta, Q_\alpha]+[\hat P_\beta, Q_\alpha]a_\beta)
-\frac{i}{2\mu}\sum_\beta[\hat P^2_\beta,\eta_\alpha]\\
&&-
\frac{i}{2\mu}\sum_\beta(a_\beta[\hat P_\beta, \eta_\alpha]+[\hat P_\beta, \eta_\alpha]a_\beta)\\
&=&\hat P_\alpha+\frac{i}{2}
(-2i a_\alpha)
-\frac{i}{2\mu}\sum_\beta[\hat P^2_\beta,\eta_\alpha]-
\frac{i}{2\mu}\sum_\beta\left(-2i\,a_\beta\frac{\partial\eta_\alpha}{\partial q_\beta}\right)\\
&=&\hat P_\alpha+a_\alpha-\frac{1}{\mu}\sum_\beta a_\beta\frac{\partial\eta_\alpha}{\partial q_\beta}-\frac{i}{2\mu}\sum[\hat P^2_\beta,\eta_\alpha].
\end{eqnarray*}
From the second and last members of this equations' chain we obtain
$-f_\alpha({\bf Q})=a_\alpha
-\frac{1}{\mu}\sum_\beta a_\beta\frac{\partial\eta_\alpha}{\partial q_\beta}-\frac{i}{2\mu}\sum[\hat P^2_\beta,\eta_\alpha]$,
which implies that $\sum_\beta[\hat P^2_\beta,\eta_\alpha]$ is a function of $\bf Q$. Therefore we have
\begin{eqnarray*}
\sum_\beta[\hat P^2_\beta,\eta_\alpha]=\phi_\alpha({\bf Q})&=&\sum_\beta(\hat P_\beta[\hat P_\beta,\eta_\alpha]+[\hat P_\beta,\eta_\alpha]\hat P_\beta)=
(-i)\sum_\beta\left(\hat P_\beta\frac{\partial\eta_\alpha}{\partial q_\beta}+\frac{\partial\eta_\alpha}{\partial q_\beta}\hat P_\beta\right)\\
&=&(-i)\sum_\beta\left(\left[\hat P_\beta,\frac{\partial\eta_\alpha}{\partial q_\beta}\right]+2\frac{\partial\eta_\alpha}{\partial q_\beta}\hat P_\beta\right)\\
&=&(-i)\sum_\beta\left((-i)\frac{\partial^2\eta_\alpha}{\partial q_\beta^2}+2\frac{\partial\eta_\alpha}{\partial q_\beta}\hat P_\beta\right).
\end{eqnarray*}
As a consequence $\sum_\beta \frac{\partial\eta_\alpha}{\partial q_\beta}\hat P_\beta$
must be a function of $\bf Q$, so that for every $\gamma$
$\sum_\beta \left[Q_\gamma,\frac{\partial\eta_\alpha}{\partial q_\beta}\hat P_\beta\right] = \nop =
\frac{\partial\eta_\alpha}{\partial q_\gamma}[Q_\gamma,\hat P_\gamma]=i\frac{\partial\eta_\alpha}{\partial q_\gamma}$;
therefore
$\frac{\partial\eta_\alpha}{\partial q_\gamma}=\nop$;
thus
$\eta_\alpha$ is a constant function. By using this result in
the equality between the second and the last members of (29)
we obtain $a_\alpha=f_\alpha$.\hfill{$\bullet$}
\section{Implying wave equations}
According to section 3.6, for a spin-0 particle the interaction described by (28), which encompasses
the electromagnetic interaction, is determined by the fact that
each operator $\eta_\alpha({\bf Q})$ appearing in the general dynamical law (27) is a real multiple
of the identity operator: $\eta_\alpha({\bf Q})=\lambda_\alpha\Id$, with $\lambda_\alpha\in{\RR}$.
Hence, according to Prop. 3.7,
$e^{i\hat G_\alpha u}=V_gU_g=e^{i\varsigma(u,{\bf Q})}U_g=
e^{i\{\eta_\alpha({\bf Q})u+o_\alpha(u,{\bf Q})\}}
U_g=
e^{i\lambda_\alpha u}e^{io_\alpha(u,{\bf Q})}U_g$,
where $o_\alpha(u,{\bf Q})$ is an operator infinitesimal of order grater than 1 in $u$ with respect to
the topology of $\mathcal H$, so that
$e^{i\hat G_\alpha u}Q_\beta^{(t)}e^{-i\hat G_\alpha u}=e^{io_\alpha(u,{\bf Q})}U_gQ_\beta^{(t)}U_g^{-1}e^{-io_\alpha(u,{\bf Q})}
=e^{io_\alpha(u,{\bf Q})}S_g[Q_\beta^{(t)}]e^{-io_\alpha(u,{\bf Q})}
=\{{\bf 1}+\omega_1(u,{\bf Q})\}S_g[Q_\beta^{(t)}]\{{\bf 1}+\omega_2(u,{\bf Q})\}$,
where $\omega_k(u,{\bf Q})$ is an operator infinitesimal of order grater than 1 in $u$.
Therefore, the $\sigma$-conversion leaves invariant the transformation properties of ${\bf Q}^{(t)}$ with
respect to galileian boosts at the first order in $u$.
\par
This result suggests that
several possible specific forms of the wave equation, i.e. of the
hamiltonian operator $H$, could be similarly determined by
this kind of invariance with respect to specific subgroups of $\mathcal G$,
also for arbitrary values of the spin.
\par
In fact, this is the
case.
In this section we shall identify which specific form the hamiltonian $H$ must take
as a consequence of the fact that the covariance properties of ${\bf Q}^{(t)}$ with respect to
specific subgroups of $\mathcal G$ are left unaltered at the first order by the $\sigma$-conversion admitted by the interaction.
In section 4.1 we address the case that such a subgroup is the subgroup of boosts, for every value of the spin.
In section 4.2 we address the task for the subgroup of spatial translations.
\subsection{Invariance under galileian boosts}
The covariance properties of ${\bf Q}^{(t)}$ with respect galileian boosts $g$ are expressed by
$S_g[Q^{(t)}_\beta]=U_gQ_\beta^{(t)}U_g^{-1}=Q^{(t)}_\beta-\delta_{\alpha\beta}ut\Id$;
therefore the equality
$$
e^{i\hat G_\alpha u}Q_\beta^{(t)}e^{-i\hat G_\alpha u}=Q_\beta^{(t)}-\delta_{\alpha\beta}ut\Id+o_1^{(t)}(u),\eqno(30)
$$
where $o_1^{(t)}(u)$ is an operator infinitesimal of order greater than 1 with respect to $u$,
is the necessary and sufficient condition in order that the $\sigma$-conversion leave unaltered
the covariance properties of ${\bf Q}^{(t)}$ with respect to
the Galileian boosts, at the first order in $u$.
\vskip.5pc\noindent
{\bf Proposition 4.1.}\;
{\sl A Q-covariant $\sigma$-conversion leaves unaltered the covariance properties of ${\bf Q}^{(t)}$ under galileian boosts at the first order in the
boosts' velocity if an only if
$$
[\eta_\alpha({\bf Q}), Q_\beta^{(t)}]=\nop\,.\eqno(31)
$$
If (31) holds, then the following relations must hold.
$$
(i)\;\;\; [\hat G_\alpha,Q_\beta^{(t)}]=i\delta_{\alpha\beta}t\;,
\quad\quad\quad\quad(ii)\;\;\;[\hat G_\alpha,\dot Q_\beta]=i\delta_{\alpha\beta}\,;\eqno(32)
$$
$$
(i)\;\;\; \mu Q_\beta^{(t)}-\hat P_\beta t=\varphi_\beta^{(t)}({\bf Q})\;,
(ii)\;\;\; \dot Q_\beta=\frac{1}{\mu}\left(\hat P_\beta+a_\beta({\bf Q})\right), \eqno(33)
$$
where $\varphi_\beta^{(t)}({\bf x})$ and $a_\beta({\bf x})=\frac{d}{dt}\varphi_\beta^{(t)}({\bf x})\mid_{t=0}$
are self-adjoint operators of ${\mathcal H}_0$.
}\vskip.4pc\noindent
{\bf Proof.}
Let $\hat U_g=e^{i\hat G_\alpha u}=V_gU_g$ be the $\sigma$-converted unitary operator associated with the galileian boost $g$, where
$V_g=e^{i\varsigma_\alpha(u,{\bf Q})}$ according to Prop. 3.7.
By starting from (30) and by expanding $e^{\pm i\varsigma_\alpha(u,{\bf Q})}$ with respect to $u$
we obtain
$$
e^{i\hat G_\alpha u}Q_\beta^{(t)}e^{-i\hat G_\alpha u}=V_gU_gQ_\beta^{(t)}U_g^{-1}V_g^{-1}
=Q_\beta^{(t)}+i[\eta_\alpha({\bf Q}),Q_\beta^{(t)}]u-\delta_{\alpha\beta}ut\Id+o_2^{(t)}(u).\eqno(34)
$$
The comparison with (30) show that such a condition holds if and only if (31) holds.
\par
By expanding $e^{\pm i\hat G_\alpha u}$ with respect to $u$ we find
$e^{i\hat G_\alpha u}Q_\beta^{(t)}e^{-i\hat G_\alpha u}=Q_\beta^{(t)}+i[\hat G_\alpha,Q_\beta^{(t)}]u+o_3^{(t)}(u)$, so that
(30) holds if and only if $i[\hat G_\alpha,Q_\beta^{(t)}]=-\delta_{\alpha\beta}t\Id$;
therefore (32) hold.
Finally,
since $\hat G_\alpha=\mu Q_\alpha$, (32.i) implies
$[\mu Q_\alpha,Q_\beta^{(t)}]=[\hat G_\alpha,Q_\beta^{(t)}]=[Q_\alpha,\hat P_\beta]t$, and then
a self-adjoint operator $\varphi_\beta^{(t)}$ of ${\mathcal H}_0$ must exists for every ${\bf x}$
such that (33) hold.
\hfill{$\bullet$}
\vskip.5pc\noindent
If we put $H_0=\frac{1}{2\mu}\sum_\gamma \left(\hat P_\gamma+a_\gamma({\bf Q})\right)^2$, then a simple calculation yields
$i[H_0,Q_\beta]=\frac{1}{\mu}\left(\hat P_\beta+a_\beta({\bf Q})\right)$.
Whenever (31) holds, Prop. 4.1 implies $i[H_0,Q_\beta]=\dot Q_\beta$, i.e.
$[H,Q_\beta]=[H_0,Q_\beta]$; therefore
$$
H=H_0+\Phi({\bf Q})=\frac{1}{2\mu}\sum_\gamma \left(\hat P_\gamma+a_\gamma({\bf Q})\right)^2+\Phi({\bf Q})
, \eqno(35)
$$
where
$\Phi({\bf x})$ is a self-adjoint operator of ${\mathcal H}_0$. Then the wave equation is
$$i\frac{\partial}{\partial t}\psi_t=\left\{\frac{1}{2\mu}\sum_\gamma \left(\hat P_\gamma+a_\gamma({\bf Q})\right)^2+\Phi({\bf Q})\right\}\psi_t.$$
\vskip.5pc
According to (35),
the dynamics of the particle is determined by the four vector valued functions $a_\alpha$, $\Phi$.
We can call them the ``fields'' which describe the effects of the interaction;
in so doing, however we have not confuse them with other notions of field involved in Quantum Physics.
Now we shall see how these fields are related to the fields
$\eta_\alpha$, $f_\alpha$ entering the general dynamical law (27).
\vskip.5pc
From (33.ii) we imply
$[\eta_\alpha({\bf Q}),\dot Q_\beta]={1\over\mu}[\eta({\bf Q}),\hat P_\beta]+{1\over\mu}[\eta_\alpha({\bf Q}),a_\beta({\bf Q})]$.
By making use of (31) we obtain
$$
{\partial\eta_\alpha\over\partial x_\beta}({\bf Q})={i\over 2}[\eta_\alpha({\bf Q}),a_\beta({\bf Q})]\,.\eqno(36.i)
$$
Now,
by replacing the form (35) of $H$ in (27) we obtain
\vskip.5pc
$\hat P_\alpha-f_\alpha({\bf Q})=i[H,\mu Q_\alpha-\eta_\alpha({\bf Q})]$\hfill{(37)}\par
\begin{eqnarray*}
&=&i\left[\frac{1}{2\mu}\sum_\beta\mu^2\dot Q_\beta^2+\Phi({\bf Q}),\mu Q_\alpha\right]
-i\left[\frac{1}{2\mu}\sum_\beta\mu^2\dot Q_\beta^2+\Phi({\bf Q}),\eta_\alpha({\bf Q})\right]\\
&=&
i\left\{\frac{1}{2}\mu\sum_\beta[\dot Q_\beta^2,\mu Q_\alpha]+[\Phi({\bf Q}),\mu Q_\alpha]\right\}+\\
&&
-i\left\{\frac{1}{2}\mu\sum_\beta[\dot Q_\beta^2,\eta_\alpha({\bf Q})]+[\Phi({\bf Q}),\eta_\alpha({\bf Q})]\right\}.
\end{eqnarray*}
By making use of (32.ii), which implies
$[\mu Q_\alpha,\dot Q_\beta^2]=2i\delta_{\alpha\beta}\dot Q_\beta$,
of (31) and of (33.ii),
we find
\begin{eqnarray*}
\hat P_\alpha-f_\alpha({\bf Q})&=&
\frac{1}{2}\mu \sum_\beta(-2i\delta_{\alpha\beta}\dot Q_\beta)+\nop-\frac{i}{2}\mu\;\nop-i[\Phi({\bf Q}),\eta_\alpha({\bf Q})]\\
&=&
\mu\dot Q_\alpha-i[\Phi({\bf Q}),\eta_\alpha({\bf Q})]=\hat P_\alpha+a_\alpha({\bf Q})-i[\Phi({\bf Q}),\eta_\alpha({\bf Q})]=\hat P_\alpha-f_\alpha({\bf Q}).
\end{eqnarray*}
Therefore we have proved that
$$
f_\alpha({\bf Q})=i[\Phi({\bf Q}),\eta_\alpha({\bf Q})]-a_\alpha({\bf Q}).\eqno(36.ii)
$$
Hence, whenever (31) holds, the fields $\eta_\alpha$ and $f_\alpha$
in the general law (27) are determined, according to (36), by the
fields $a_\alpha$, $\Phi$.
\vskip.5pc
For the particular case of a spin-0 particle we can show the following further characterization.
\vskip.5pc\noindent
{\bf Proposition 4.2.}
{\sl
In the simplest quantum Theory of an interacting particle, corresponding to the case ${\mathcal H}_0=\CC$
in (21), the Q-covariant $\sigma$-conversions for which $\eta_\alpha({\bf Q})=$\,constant are those and only those
which leave unaltered the covariant properties of ${\bf Q}^{(t)}$ with respect to the Galileian
boosts $g\in\mathcal G$, at the first order in the boost's velocity.}
\vskip.4pc\noindent
{\bf Proof.}
If $\eta_\alpha=$constant then (31) holds, of course. Therefore, in order to prove the proposition, it is sufficient
to prove the inverse implication.
Hence we suppose that (31) holds. It implies the condition $[\eta_\alpha({\bf Q}),Q_\alpha]=\nop$.
On the other hand, (33.i) implies $Q_\beta^{(t)}={t\over\mu}\left(\varphi_\beta^{(t)}({\bf Q})+\hat P_\beta\right)$,
which replaced in (31) yields
$[\eta_\alpha({\bf Q}),\hat P_\beta]=\nop$;
therefore $\eta_\alpha({\bf Q})$ is a constant operator $\lambda_\alpha\Id$.\hfill{$\bullet$}
\subsection{Invariance under spatial translations}
Let us now suppose that the interaction admits a Q-covariant $\sigma$-conversion such that if $\hat U_g=e^{-i\hat P_\alpha a}$ then
$$
e^{-i\hat P_\alpha a}Q_\beta^{(t)}e^{i\hat P_\alpha a}=Q_\beta^{(t)}-\delta_{\alpha\beta}+o_1^{(t)}(a),\eqno(38)
$$
where $o_1^{(t)}(a)$ is an infinetisimal operator of order greater than 1 in $a$.
In fact, we are supposing that
the interaction leaves unaltered the covariance properties of ${\bf Q}^{(t)}$ with respect to spatial translations
at the first order in the translation parameter $a$. Now, by expanding $e^{-i\hat P_\alpha a}$ with respect to
the translation parameter $a$,
(38) yields
$$
(i)\;\;\;[Q_\beta^{(t)},\hat P_\alpha]=i\delta_{\alpha\beta}\quad\hbox{ which implies }\quad (ii)\;\;\; [\dot Q_\beta,\hat P_\alpha]=\nop.
\eqno(39)
$$
Therefore we can state that
$$
\dot Q_\beta=v_\beta(\hat{\bf P}),
\eqno(40)
$$
where $v_\beta({\bf p})$ is a self-adjoint operator of ${\mathcal H}_0$.
Since $[Q_\alpha,v_\beta(\hat{\bf P})]=i\frac{\partial v_\beta}{\partial p_\alpha}(\hat{\bf P})$, by making use of the Jacobi identity
for $\left[Q_\alpha,[H,Q_\beta]\right]$ we obtain
$$
i\frac{\partial v_\beta}{\partial p_\alpha}(\hat{\bf P})=[Q_\alpha,\dot Q_\beta]
=i\left[Q_\alpha,[H,Q_\beta]\right]
=[Q_\beta,\dot Q_\alpha]
=i\frac{\partial v_\alpha}{\partial p_\beta}(\hat{\bf P}).
$$
This equality shows that ${\bf v}({\bf p})=\left(v_1({\bf p}),v_2({\bf p}),v_3({\bf p})\right)$ is an irrotational field; hence a function
$F$ of $\bf p$ exists such that $v_\alpha({\bf p})=\frac{\partial F}{\partial p_\alpha}({\bf p})$, where $F({\bf p})$ is
a self-adjoint operator of ${\mathcal H}_0$.
Therefore we can establish the following equalities.
$$
\dot Q_\alpha=v_\alpha(\hat{\bf P})=\frac{\partial F}{\partial p_\alpha}(\hat{\bf P})=i[F(\hat{\bf P}),Q_\alpha]=i[H,Q_\alpha].\eqno(41)
$$
The last equation implies that a function $\Psi$ of $\bf x$ exists such that
$H-F(\hat{\bf P})=\Psi(\hat{\bf Q})$, i.e.
$$
H=F(\hat{\bf P})+\Psi({\bf Q}),\eqno(42)
$$
where $\Psi({\bf x})$ is a self-adjoint operator of ${\mathcal H}_0$. Then the wave equation is
$$i\frac{\partial}{\partial t}\psi_t=\left\{F(\hat{\bf P})+\Psi({\bf Q})\right\}\psi_t.$$
\subsection{Invariance under both}
Let us suppose that the interaction admits a $\sigma$-conversion that leaves unaltered the
covariance properties of ${\bf Q}^{(t)}$ under both subgroups of boosts and of spatial translations.
Accordingly, the following equality holds
\begin{eqnarray*}
H&=&F(\hat{\bf P})+\Psi({\bf Q})=\frac{1}{2\mu}\sum_\gamma \left(\hat P_\gamma+a_\gamma({\bf Q})\right)^2+\Phi({\bf Q})\\
&=&\frac{1}{2\mu}\sum_\gamma \left({\hat P}_\gamma^2+a_\gamma({\bf Q})\hat P_\gamma+\hat P_\gamma a_\gamma({\bf Q})+a_\gamma^2({\bf Q})\right)
+\Phi({\bf Q}).
\end{eqnarray*}
Since
$a_\gamma({\bf Q})\hat P_\gamma+\hat P_\gamma a_\gamma({\bf Q})=
[a_\gamma({\bf Q}),\hat P_\gamma]+2a_\gamma({\bf Q})\hat P_\gamma=
i\frac{\partial a_\gamma}{\partial x_\gamma}({\bf Q})+2a_\gamma({\bf Q})\hat P_\gamma$
the equality above implies
$$
\frac{1}{2\mu}\sum_\beta\hat P_\beta a_\beta({\bf Q})=\left(F(\hat{\bf P})-\frac{1}{2\mu}\sum_\beta {\hat P}_\beta^2\right)
+\Psi({\bf Q})-\frac{i}{2\mu}\sum_\beta\frac{\partial a_\beta}{\partial x_\beta}({\bf Q})-\Phi({\bf Q})-\sum_\beta a_\beta^2({\bf Q})\,.
$$
Then
$$
\frac{1}{2\mu}\sum_\beta\hat P_\beta a_\beta({\bf Q})=F_1(\hat{\bf P})+F_2({\bf Q}),
$$
where $F_1(\hat{\bf P})=\left(F(\hat{\bf P})-\frac{1}{2\mu}\sum_\beta {\hat P}_\beta^2\right)$ and
$F_2({\bf Q})=\Psi({\bf Q})-\frac{i}{2\mu}\sum_\beta\frac{\partial a_\beta}{\partial x_\beta}({\bf Q})-\Phi({\bf Q})-\sum_\beta a_\beta^2({\bf Q})$.
Therefore
$$
\left[Q_\gamma,\frac{1}{2\mu}\sum_\beta\hat P_\beta a_\beta({\bf Q})\right]=
\frac{i}{2\mu}a_\gamma({\bf Q})=\frac{\partial f_1}{\partial x_\gamma}(\hat {\bf P}).
$$
Then
$$
[\hat P_\alpha,a_\gamma({\bf Q}]=\frac{\partial a_\gamma}{\partial x_\alpha}({\bf Q})=
-2i\mu \left[\hat P_\alpha,\frac{\partial f_1}{\partial x_\gamma}(\hat {\bf P})\right]=\nop.
$$
Therefore, $a_\gamma({\bf Q})$ is an operator that acts as follows
$$
\left(a_\gamma({\bf Q})\psi\right)({\bf x})=\hat a_\gamma\psi({\bf x}),
$$
where $\hat a_\gamma$ is an operator of ${\mathcal H}_0$ which does not depend on $\bf x$.
\par
Thus, if (31) and (38) hold, then
$H=\frac{1}{2\mu}\sum_\gamma(\hat P_\gamma+\hat a_\gamma)^2+ \Phi({\bf Q})$,
and the wave equation is
$$i\frac{\partial}{\partial t}\psi_t=
\left\{\frac{1}{2\mu}\sum_\gamma(\hat P_\gamma+\hat a_\gamma)^2+ \Phi({\bf Q})\right\}\psi_t\,.,\eqno(43)$$
In the spin-0 case, it is straightforward to show that if the $\sigma$-conversion leaves unaltered also the covariance
properties of ${\bf Q}^{(t)}$ with respect to the rotations subgroup, then $a_\gamma=0$.
\subsection{Conclusive remarks}
Our work has been successful in deriving the known forms (35) and (43) of the non relativistic wave equation
of an interacting particle, by implying them through a deductive development. However, the present approach
does not exclude the possibility of wave equations, and hence of interactions, different from the known ones.
\par
In fact, the existence of interactions besides those described by equations (35) and (43) is manifestly required
by the phenomenological reality. It is sufficient to recall that (35) describes, in its classical limit, only
{\sl approximately} the motion of a charged particle in an electromagnetic field; the approximation consists in
neglecting the effect of the radiation caused by the acceleration of the charge on the motion of the particle.
Actually, (35) can be considered valid only for particles {\sl slowly} accelerated by the electromagnetic field.
Since the occurrence of phenomena with {\sl strongly} accelerated particles cannot be
excluded, a non approximated wave equation must be different from (35) and from (43). According
to our approach this implies that the corresponding $\sigma$-conversion
cannot leave unaltered the covariance properties of
${\bf Q}^{(t)}$ with respect to the subgroup of boosts.
\par
Therefore our development opens to the possibility for interactions different form those described by (35) and (43).
The present work has a general character, and these possibilities have
not been specifically explored; however, the present theoretical framework allows for such an investigation.
This possibility is precluded by the
other practiced methods for quantizing the interaction;
for instance the methods based on the gauge principle -- without entering Quantum Field Theories --
immediately lead to wave equations of the form (35).
On the other hand, the method of canonical quantization is constitutionally constrained to the wave
equations implied by the classical equations.
\par
The author is convinced that interesting results of the approach can be found by extending it
to a relativistic theory. Such an extension is in progress.


\end{document}